\documentclass[]{aa} 

\usepackage{natbib}
\bibpunct{(}{)}{;}{a}{}{,} % to follow the A&A style
\usepackage{epsfig}

\newcommand{\beq}{\begin{equation}} 
\newcommand{\eeq}{\end{equation}} 
\newcommand{\beqa}{\begin{eqnarray}} 
\newcommand{\eeqa}{\end{eqnarray}} 
\newcommand{\bea}{\begin{array}} 
\newcommand{\ea}{\end{array}} 

\newcommand{\dd}{{\rm d}}
\newcommand{\pl}{\partial}
 
\newcommand{\lag}{\langle} 
\newcommand{\rag}{\rangle} 

\newcommand{\ii}{{\rm i}}

\newcommand{\Om}{\Omega_{\rm m}} 
\newcommand{\rhob}{\overline{\rho}} 
\newcommand{\xib}{\overline{\xi}}
\newcommand{\kpar}{k_{\parallel}}

\newcommand{\ve}{{\bf e}}
\newcommand{\vk}{{\bf k}}
\newcommand{\vK}{{\bf K}}
\newcommand{\vkperp}{{\bf k}_{\perp}}
\newcommand{\vq}{{\bf q}}
\newcommand{\vs}{{\bf s}}
\newcommand{\vv}{{\bf v}}
\newcommand{\vw}{{\bf w}}
\newcommand{\vx}{{\bf x}}
\newcommand{\vPsi}{{\bf \Psi}}

\newcommand{\tdelta}{{\tilde{\delta}}}
\newcommand{\tvPsi}{{\tilde{\bf \Psi}}}

\newcommand{\cF}{{\cal{F}}}
 \newcommand{\cG}{{\cal{G}}}

\newcommand{\erfc}{{\rm erfc}}
\newcommand{\Psc}{P_{\rm s.c.}}
\newcommand{\Psticky}{P_{\rm sticky}}
\newcommand{\PZel}{P_{\rm Zel}}
\newcommand{\Real}{{\rm Re}}
\newcommand{\PZelpar}{P^s_{\parallel \rm Zel}}
\newcommand{\Pscpar}{P^s_{\parallel \rm s.c.}}

\begin{document}

\title{Impact of shell crossing and scope of perturbative approaches in real
and redshift space.} 
\author{P. Valageas}   
\institute{Institut de Physique Th\'eorique, CEA Saclay, 91191 Gif-sur-Yvette, 
France}  
\date{Received / Accepted } 
 
\abstract
{}
{We study the effect of nonperturbative corrections associated with
the behavior of particles after shell crossing on the matter power spectrum.
We compare their amplitude with the perturbative terms that can be obtained
within the fluid description of the system, to estimate the range of scales where
such perturbative approaches are relevant.}
{We use the simple  Zeldovich dynamics as a benchmark, as it allows the exact
computation of the full nonlinear power spectrum and of perturbative terms at
all orders. Then, we introduce a ``sticky model'' that coincides with the Zeldovich
dynamics before shell crossing but shows a different behavior afterwards. Thus,
their power spectra only differ in their nonperturbative terms.
We consider both the real-space and redshift-space power spectra.}
{We find that the potential of perturbative schemes is greater at higher redshift
for a $\Lambda$CDM cosmology. For the real-space power spectrum, one can go
up to order $66$ of perturbation theory at $z=3$, and to order $9$ at $z=0$,
before the nonperturbative correction surpasses
the perturbative correction of that order. This allows us to increase the upper bound
on $k$ where systematic theoretical predictions may be obtained by perturbative
schemes, beyond the linear regime, by a factor $\sim 26$ at $z=3$ and
$\sim 6.5$ at $z=0$. This provides a strong motivation to study perturbative
resummation schemes, especially at high redshifts $z \geq 1$.

In the context of cosmological reconstruction methods,
the Monge-Amp\`ere-Kantorovich scheme appears to be close to optimal at $z=0$.
There also seems to be little room for improvement over current reconstruction
methods of the baryon acoustic oscillations at $z=0$.
This can be understood from the small number of perturbative terms that are
relevant at $z=0$, before nonperturbative corrections dominate.

We also point out that the rise of the power spectrum on the transition
scale to the nonlinear regime strongly depends on the behavior of the system
after shell crossing.

We find similar results for the redshift-space power spectrum, with characteristic
wavenumbers that are shifted to lower values as redshift-space distortions amplify
higher order terms of the perturbative expansions while decreasing the resummed
nonlinear power at high $k$.}
{}

\keywords{gravitation; cosmology: theory -- large-scale structure of Universe}

\maketitle

\section{Introduction} 
\label{Introduction}

The growth of large-scale structures in the Universe through the amplification
of small primordial fluctuations by gravitational instability is a key ingredient
of modern cosmology \citep{Peebles1980}, and it can be used to constrain 
cosmological parameters through the dependence of the matter power spectrum
on scale and redshift. 
On very large scales or at high redshifts, where the amplitude of the density
fluctuations is small, it is sufficient to use linear theory, whereas on small scales,
in the highly nonlinear regime, one must use numerical simulations or
phenomenological models, such as the halo model \citep{Cooray2002}, which are
also calibrated on simulations.
In the weakly nonlinear regime one expects perturbative approaches to provide
a useful tool, as they allow going beyond linear theory in a systematic and controlled
fashion. Several observational probes, such as weak lensing surveys
\citep{Massey2007,Munshi2008} or measures of acoustic baryonic oscillations
\citep{Eisenstein1998,Eisenstein2005}, are mostly sensitive to these intermediate
scales, and to meet the accuracy of future observations we need
a theoretical accuracy of about $1\%$. Phenomenological models
typically have an accuracy of $10\%$ in this range, while numerical simulations 
may suffer from finite resolution and finite size effects and require a long
computational time if we need to obtain the power spectra over a fine grid of
cosmological parameters.

This has led to a renewed interest in perturbative approaches, as
it may be possible to improve over the standard perturbation theory 
\citep{Goroff1986,Bernardeau2002} by using resummation schemes that allow
systematic partial resummations of higher order terms .
Thus, \citet{Crocce2006a,Crocce2006b} present a partial resummation
of the diagrammatic series associated with the response function (propagator),
within a high-$k$ limit, which provides improved predictions for the density power
spectrum \citep{Crocce2008}. On the other hand, \citet{Valageas2007a}
describes a path-integral formalism that allows applying the tools of field theory,
such as large-$N$ expansions,
to compute the power spectrum and higher order statistics like the
bispectrum \citep{Valageas2008}. One of these  large-$N$ expansions was recovered
by \citet{Taruya2008}, as a ``closure theory'' where one closes the hierarchy
of equations satisfied by the many body correlations at the third order, following
the ``direct interaction approximation'' introduced in hydrodynamics
\citep{Kraichnan1959}. This also improves the predictions for the power spectrum
on the scales probed by the baryonic acoustic oscillations
\citep{Taruya2009,Valageas2010a}.
Other approaches have been proposed by \citet{Matarrese2007}, using the dependence
on a running high-$k$ cutoff, by \citet{Pietroni2008}, using a truncation of the hierarchy
satisfied by the many-body correlations, by \citet{Matsubara2008}, within a Lagrangian
framework, and by \citet{McDonald2007}, using a renormalization group technique.
Most\footnote{Two exceptions are the formalism developed in \citet{Valageas2004},
which applies to the Vlasov equation, and the Lagrangian approach of
\citet{Matsubara2008}, which however is not valid beyond shell crossing.}
of these approaches start from the fluid description of the system,
where the density and velocity fields obey hydrodynamical equations of motion.
This corresponds to a single-stream approximation that neglects shell crossing.
Then, the domain of validity of most such perturbative
schemes is limited to wavenumbers where shell-crossing effects are negligible,
even if we could sum all perturbative terms. 

This problem with the impact of shell crossing also appears
in the context of cosmological reconstruction, where one attempts to follow
the matter distribution observed in a given galaxy survey back in time
\citep{Peebles1989}. Thus, an efficient algorithm for building such a reconstruction
(which can then be used  to estimate the velocity field) is provided by the
Monge-Amp\`ere-Kantorovich method, which neglects multistreaming
\citep{Brenier2003,Mohayaee2006}. 
Similar methods are used to reconstruct the baryon acoustic
oscillations (BAO) of the density power spectrum, in order to improve the accuracy of
cosmological distance measurements and tighten the constraints on cosmological
parameters such as the amount and evolution of dark energy
\citep{Eisenstein2007,Seo2010}.
Then, by comparing the amplitude of
nonperturbative corrections with the perturbative terms obtained within
the fluid description, one can estimate the scale down to which these reconstruction
schemes can be used. 

Thus, to estimate the potential of such approaches, it is necessary to evaluate
the effect of shell crossing on the matter power spectrum. 
This question has already been investigated by \citet{Afshordi2007} by
comparing the phenomenological halo model with a modified variant where halos
are collapsed to pointlike masses. In this paper we revisit this problem in a more
systematic fashion, within the framework of the Zeldovich
dynamics. Thus, we compare the Zeldovich dynamics with a second model (named
the ``sticky model'' in the following),
which only differs after shell crossing. Therefore, both power spectra have the
same perturbative expansions and only differ by nonperturbative terms.
Taking advantage of our being able to explicitly compute the full nonlinear
spectra, as well as perturbative terms at all orders and these nonperturbative
corrections, we can compare their respective amplitudes in detail.
This allows a more detailed discussion of the importance of shell-crossing
effects and of the scope of perturbative approaches. This also enables us to
distinguish the dependence on the behavior of the dynamics after shell crossing
of the rise of the density power spectrum on mildly nonlinear scales, in the
intermediate regime where the logarithmic power goes from $\Delta^2(k) \sim 1$
to $\Delta^2(k) \sim 100$. In addition, we can perform the same analysis for
the redshift-space power spectrum, which is actually the quantity most directly
observed in galaxy surveys.

This article is organized as follows. We first recall in Sect.~\ref{Non-linear-Zeldovich}
the nonlinear real-space power spectrum associated with the Zeldovich dynamics, and its
perturbative expansions in Sects.~\ref{Standard-perturbative} and
\ref{Renormalized-perturbative}. We present our ``sticky model'', which only differs
from the Zeldovich dynamics after shell crossing, in Sect.~\ref{nonperturbative},
and we give the associated nonperturbative correction. Then, we describe
the numerical results obtained for a $\Lambda$CDM cosmology in Sect.~\ref{Numerical},
comparing the various perturbative and nonperturbative terms.
Finally, we extend our analysis to the redshift-space power spectrum in
Sect.~\ref{Redshift-space-power-spectrum}, focusing on wavenumbers that are
aligned with the line of sight.
We conclude in Sect.~\ref{Conclusion}.

\section{Computation of the density power spectrum in real space}
\label{Computation}

\subsection{Nonlinear Zeldovich power spectrum}
\label{Non-linear-Zeldovich}

As is well known, the Zeldovich approximation \citep{Zeldovich1970}
sets the Eulerian position, $\vx(\vq,t)$, of the particle of Lagrangian
coordinate $\vq$, equal to the position given by the linear displacement
field, $\vPsi_L(\vq,t)$,
\beq
\vx(\vq,t) = \vq + \vPsi_L(\vq,t) \;\;\; \mbox{with} \;\;\;
\nabla_{\vq} \cdot \vPsi_L= -\delta_L(\vq,t) ,
\label{Zeldef}
\eeq
where $\delta_L(\vq,t)$ is the linear growing mode of the density contrast,
which is defined by
\beq
\delta(\vx,t) = \frac{\rho(\vx,t)-\rhob}{\rhob} .
\label{deltadef}
\eeq
Here $\rhob$ is the mean matter density of the Universe, we work in comoving
coordinates and as usual we only consider the linear growing mode.
It is well known \citep{Schneider1995,Taylor1996}
that the explicit expression of the matter power spectrum
can be derived from Eq.(\ref{Zeldef}) by using the conservation of matter, which
reads as
\beq
\rho(\vx) \, \dd\vx = \rhob \, \dd\vq \;\;\; \mbox{whence} \;\;\; 
1+\delta(\vx) = \left|\det\left(\frac{\pl\vx}{\pl\vq}\right)\right|^{-1} .
\label{rhox}
\eeq
For an arbitrary displacement field $\vPsi$, this also reads as
\beq
1+\delta(\vx) = \int\dd\vq \; \delta_D[\vx-\vq-\vPsi(\vq)] ,
\label{deltax}
\eeq
where $\delta_D$ is the Dirac distribution, and this yields in Fourier
space (for $k\neq 0$, that is, disregarding a term $\delta_D(\vk)$):
\beq
\tdelta(\vk) = \int\frac{\dd\vx}{(2\pi)^3} \, e^{-\ii\vk\cdot\vx} \, \delta(\vx)
= \int\frac{\dd\vq}{(2\pi)^3} \, e^{-\ii\vk\cdot(\vq+\vPsi)} .
\label{tdelta}
\eeq
Defining the density power spectrum as
\beq
\lag \tdelta(\vk_1) \tdelta(\vk_2) \rag = \delta_D(\vk_1+\vk_2) P(k_1) ,
\label{Pkdef}
\eeq
we obtain from Eq.(\ref{tdelta}), using statistical homogeneity,
\beq
P(k) = \int\frac{\dd\vq}{(2\pi)^3} \, \lag e^{\ii \vk \cdot [\vx(\vq)-\vx(0)]} \rag .
\label{Pkxq}
\eeq
Equation (\ref{Pkxq}) is quite general since we have not used the
Zeldovich approximation (\ref{Zeldef}) yet. It shows how the density
power spectrum is related to the statistical properties of the displacement
field, for any mapping $\vq\mapsto\vx$ (which can include some shell-crossing,
as in the Zeldovich case). 
The great simplification provided
by the Zeldovich approximation (\ref{Zeldef}) is that in this case the
quantity $\vx(\vq)-\vx(0)$ is a Gaussian random variable, so that the
mean can be computed at once as
\beq
P(k) = \int\frac{\dd\vq}{(2\pi)^3} \, e^{\ii\vk\cdot\vq} \,
e^{-\frac{1}{2} \lag (\vk \cdot [\vPsi_L(\vq)-\vPsi_L(0)])^2 \rag} .
\label{Pksq}
\eeq
Next, using the second relation (\ref{Zeldef}) to compute the average
$\lag (\vk \cdot [\vPsi_L(\vq)-\vPsi_L(0)])^2 \rag$, we obtain the explicit expression
\beq
P(k) = \int\frac{\dd\vq}{(2\pi)^3} \, e^{\ii\vk\cdot\vq} \,
e^{-\int \dd\vw \,
[1-\cos(\vw\cdot\vq)] \, \frac{(\vk\cdot\vw)^2}{w^4} \, P_L(w)} .
\label{Pkw}
\eeq
It is convenient to perform the integration over the angles of the
wavenumber $\vw$ by expanding $e^{\ii \vw\cdot\vq}$ over spherical
harmonics \citep{Schneider1995},
\beq
\int \dd\vw \,
e^{\ii \vw\cdot\vq} \, \frac{(\vk\cdot\vw)^2}{w^4} \, P_L(w)
= k^2 I_0(q) + k^2 (1-3\mu^2) I_2(q) ,
\eeq
where $\mu=(\vk\cdot\vq)/(k q)$, and we introduced
\beq
I_{\ell}(q) = \frac{4\pi}{3} \int_0^{\infty} \dd w \, P_L(w) \; j_{\ell}(qw) ,
\label{Inqdef}
\eeq
where $j_{\ell}$ is the spherical Bessel function of order $\ell$.
In particular, the variance $\sigma_v^2$ of the one-dimensional
displacement field (or of the linear velocity field, up to a time-dependent
multiplicative factor), reads as
\beq
\sigma_v^2 = \frac{1}{3} \lag |\vPsi_L|^2\rag = I_0(0) .
\label{sigmav}
\eeq
Therefore, the expression (\ref{Pkw}) also writes as
\beq
P(k) = \int\frac{\dd\vq}{(2\pi)^3} \, \cos(kq\mu) \,
e^{-k^2[\sigma_v^2-I_0(q)-(1-3\mu^2)I_2(q)]} .
\label{PkIn}
\eeq
Following \cite{Schneider1995}, we can perform the integration over the
angles of $\vq$ by expanding part of the exponential, and using the
property
\beq
\int_0^1\dd\mu \, \cos(kq\mu) \, (1-\mu^2)^{\ell} = \ell !
\left(\frac{2}{kq}\right)^{\ell} j_{\ell}(kq) ,
\label{int-mu}
\eeq
which gives
\beqa
P(k) \!\! & = \! & \!\! \int\! \frac{\dd q \, q^2}{2\pi^2} \,
e^{-k^2[\sigma_v^2-I_0(q)+2I_2(q)]}  \sum_{\ell=0}^{\infty}
\left(\frac{6kI_2(q)}{q}\right)^{\!\ell}
 j_{\ell}(kq) .  \nonumber \\
&&
\label{Pkjn}
\eeqa

\subsection{Standard perturbative expansion}
\label{Standard-perturbative}

In the standard perturbative approach \citep{Goroff1986,Bernardeau2002},
one writes the nonlinear density contrast as a series over powers of the
linear growing mode,
\beq
\tdelta(\vk) = \sum_{n=1}^{\infty} \tdelta^{(n)}(\vk)  \;\;\; \mbox{with}
\;\;\; \tdelta^{(n)}(k) \propto (\tdelta_L)^{n} ;
\label{deltal}
\eeq
that is,
\beqa
\tdelta^{(n)}(\vk)  & = & \int \dd\vw_1 .. \dd\vw_{n} \, 
\delta_D(\vw_1+..+\vw_{n}-\vk) \, F_{n}(\vw_1, .. ,\vw_{n}) \nonumber \\
&& \times \, \tdelta_L(\vw_1) .. \tdelta_L(\vw_{n}) .
\label{Fndef}
\eeqa
Substituting this expansion (and the one associated with the velocity field)
into the equations of motion one obtains a recursion relation for the kernels
$F_{n}$, which allows terms of increasing order to be computed in a sequential
manner. As is well known, for the Zeldovich dynamics, where $\vPsi=\vPsi_L$,
we do need to follow this route, since by expanding the exponential (\ref{tdelta})
over $\vPsi_L$, and using $\tvPsi_L(\vk)= \ii (\vk/k^2) \tdelta_L(\vk)$, we obtain
at once all terms,
\beq
F_{n}(\vw_1, .. ,\vw_{n}) = \frac{1}{n !} \, \frac{\vk\cdot\vw_1}{w_1^2} \,
... \, \frac{\vk\cdot\vw_{n}}{w_n^2} .
\label{FnZel}
\eeq
Then, substituting the expansion (\ref{deltal}) into the definition (\ref{Pkdef})
of the power spectrum and taking the Gaussian average, one obtains the
standard perturbative series
\beq
P(k) = \sum_{n=1}^{\infty} P^{(n)}(k) \;\; \mbox{with}
\;\; P^{(n)}(k) \propto (P_L)^{n} .
\label{Pstd}
\eeq
In particular, the two lowest order terms are
\beq
P^{(1)}(k)= P_L(k) , \;\; P^{(2)}(k)= P_{22}(k)+P_{13}(k) ,
\eeq
where $P_{22}$ and $P_{13}$ arise from terms of the form
$\lag \tdelta^{(2)} \tdelta^{(2)}\rag$ and $\lag \tdelta^{(1)} \tdelta^{(3)}\rag$, with
\beqa
P_{22}(k) & = & \int\dd\vw_1\dd\vw_2 \, \delta_D(\vw_1+\vw_2-\vk) \,
\frac{(\vk\cdot\vw_1)^2(\vk\cdot\vw_2)^2}{2w_1^4w_2^4} \nonumber \\
&& \times P_L(w_1) P_L(w_2) ,
\label{P22}
\eeqa
and
\beq
P_{13}(k) = - k^2\sigma_v^2 \, P_L(k) .
\label{P13}
\eeq
In fact, for the Zeldovich dynamics it is not necessary to use this procedure,
since by expanding the exponential in the exact result (\ref{Pkw}) or (\ref{Pkjn})
we obtain all terms at once,
\beqa
P^{(n)}(k) & = & \int_0^{\infty}\frac{\dd q \, q^2}{2\pi^2} \, \sum_{p=0}^{n} \,
\frac{1}{p!} \, [k^2(I_0(q)-2I_2(q)-\sigma_v^2)]^p \nonumber \\
&& \times \left(\frac{6kI_2(q)}{q}\right)^{n-p} \, j_{n-p}(kq) .
\label{Pnstd}
\eeqa

\subsection{Renormalized perturbative expansion}
\label{Renormalized-perturbative}

As pointed out by \cite{Crocce2006a}, it is useful to keep the $\vq$-independent
exponential term $e^{-k^2\sigma_v^2}$ in Eq.(\ref{PkIn}) and to define
a ``renormalized'' perturbative expansion
\beq
P(k) =  e^{-k^2\sigma_v^2} \sum_{n=1}^{\infty} P_{\sigma_v}^{(n)}(k)
\;\; \mbox{with}
\;\; P_{\sigma_v}^{(n)}(k) \propto (P_L)^{n} .
\label{Psigv}
\eeq
Indeed, while the standard perturbative terms $P^{(n)}$ grow increasingly
fast at high $k$ and show large cancellations, the ``renormalized'' terms
$e^{-k^2\sigma_v^2} P_{\sigma_v}^{(n)}$ are positive and show a Gaussian
decay at high $k$.
Thus, each term peaks on a well-defined range of $k$ and one can clearly see
the contribution of diagrams of a given order to the full nonlinear power spectrum.
In fact, as noticed in \cite{Crocce2006a}, these terms also write as
\beqa
P_{\sigma_v}^{(n)}(k) & = & n! \int \dd\vw_1 .. \dd\vw_{n} \,
\delta_D(\vw_1+..+\vw_{n}-\vk)  \nonumber \\
&& \times \, F_{n}(\vw_1, .. ,\vw_{n})^2 \, P_L(w_1) .. P_L(w_{n}) .
\label{Pn-sigvFn}
\eeqa
This can be seen at once by expanding the expression (\ref{Pkw})
and recognizing the square of the kernel $F_n$ given in Eq.(\ref{FnZel}).
Thus, each new term $P_{\sigma_v}^{(n)}$ is associated with the kernel 
$F_n$ that couples $n$ linear modes, so that one can follow the contribution
of higher order mode couplings.
Again, from Eq.(\ref{Pkjn}) we obtain all terms at once,
\beqa
P_{\sigma_v}^{(n)}(k) & = & \int_0^{\infty}\frac{\dd q \, q^2}{2\pi^2} \,
\sum_{p=0}^{n} \, \frac{1}{p!} \, [k^2(I_0(q)-2I_2(q))]^p \nonumber \\
&& \times \left(\frac{6kI_2(q)}{q}\right)^{n-p} \, j_{n-p}(kq) ,
\label{Pn-sigv}
\eeqa
while the two lowest order terms are
\beq
P_{\sigma_v}^{(1)}(k) = P_L(k) , \;\;\; P_{\sigma_v}^{(2)}(k) = P_{22}(k) ,
\eeq
where $P_{22}$ was given in Eq.(\ref{P22}).

For numerical purposes, it is convenient to obtain the standard perturbative
terms (\ref{Pstd}) from the renormalized ones (\ref{Psigv}). Expanding the
prefactor in Eq.(\ref{Psigv}) gives
\beq
P^{(n)}(k) = \sum_{p=0}^{n-1} \frac{1}{p!} \, (-k^2\sigma_v^2)^p \,
P_{\sigma_v}^{(n-p)}(k)  .
\eeq
On the other hand, to compute the full nonlinear power spectrum (\ref{Pkjn})
it is convenient to subtract the two terms
$e^{-k^2\sigma_v^2} [P_{\sigma_v}^{(1)}(k)+P_{\sigma_v}^{(2)}(k)]$,
so as to avoid integrating over badly behaving terms for $q\rightarrow\infty$.

\subsection{Nonperturbative correction}
\label{nonperturbative}

The expressions recalled in the previous sections give the nonlinear power
spectrum and its perturbative expansions for the usual Zeldovich dynamics,
where particles follow the linear trajectories (\ref{Zeldef}). This includes
shell crossing and leads to a decay of the nonlinear power spectrum
at high $k$, as particles freely escape to infinity, and these random trajectories
erase small-scale features
\citep{Schneider1995,Taylor1996,Valageas2007b,BernardeauVal2010a}.
However, because of the simple nature of the Zeldovich dynamics, the
full nonlinear result (\ref{PkIn}) can be obtained by
resumming\footnote{The radius of convergence of such
series can be finite (e.g., for a power-law linear spectrum $P_L(k) \propto k^n$
with $n=-2$) or zero (if $n<-2$), see \citet{Valageas2007b}.}
the perturbative expansions (\ref{Pstd}) or (\ref{Psigv}). The latter can be obtained
from the usual hydrodynamical equations of motion in Eulerian space,
which actually break down at shell crossing.

Our goal in this work is to compare the perturbative terms (\ref{Pstd}) 
and (\ref{Psigv}) with the nonperturbative terms that are associated with the
physics that takes place after shell crossing.
Within the framework of the Zeldovich dynamics considered in this article,
this means that we wish to compare the previous results with those that
would be obtained for a second dynamics, which coincides with the
Zeldovich dynamics until shell crossing.
Then, the difference between both power spectra would give us the amplitude
of the correction due to shell-crossing effects, which are generically 
nonperturbative.

To introduce this second model, let us first recall that the linear growing mode
of the displacement field, $\vPsi_L(\vq)$, and the associated peculiar
velocity field, $\vv_L(\vq) \propto \vPsi_L(\vq)$, are curlfree \citep{Peebles1980},
and are derived from a velocity potential, $\vv_L=-\nabla\cdot\chi_L$.
Using the Poisson equation we can see that this potential is equal to the
linear gravitational potential $\Phi_L$, up to a time-dependent factor
\citep{Peebles1980,Vergassola1994}. 
In particular, the Lagrangian-space to Eulerian-space mapping defined
by the Zeldovich dynamics (\ref{Zeldef}) derives from a Lagrangian
potential $\varphi_L(\vq)$,
\beq
\vx(\vq,t) = \frac{\pl\varphi_L}{\pl\vq} ,
\label{xq}
\eeq
with
\beq
\varphi_L(\vq,t) = \frac{|\vq|^2}{2} -  \frac{a(t)}{4\pi \cG \rhob} \, \Phi_L(\vq,t) ,
\label{phiLdef}
\eeq
where $\Phi_L$ is the linear gravitational potential, which obeys 
the Poisson equation,
\beq
\Delta_{\vq} \Phi_L = \frac{4\pi \cG \rhob}{a(t)} \, \delta_L .
\label{Poisson}
\eeq
Here $\cG$ is Newton's constant and $a(t)$ the scale factor.
Then, from Eq.(\ref{rhox}) the nonlinear density contrast is given by the
Hessian determinant of $\varphi_L$,
\beq
1+\delta(\vx) =  \left|\det\left(\frac{\pl\vx}{\pl\vq}\right)\right|^{-1} 
= \left|\det\left(\frac{\pl^2\varphi_L}{\pl q_i \pl q_j}\right)\right|^{-1} .
\label{Hessian}
\eeq
At early times, $t\rightarrow 0$, (and if there is not too much power on small scales),
the Lagrangian potential $\varphi_L$ is dominated by the first term in
Eq.(\ref{phiLdef}) and $\vx(\vq,t) \rightarrow \vq$. Then, the Hessian matrix
$(\pl^2\varphi_L/\pl q_i \pl q_j)$ is definite positive and goes to the identity
matrix, whence $\delta\rightarrow 0$. As time increases
and structures form, the Lagrangian potential becomes increasingly sensitive
to the fluctuations in the linear gravitational potential, and the Hessian
determinant deviates from unity. However, the eigenvalues remain strictly positive
until shell crossing, which means that the  Lagrangian potential $\varphi_L$
remains a strictly convex function. 
Then, at shell crossing one eigenvalue
goes through zero and becomes negative (generically the collapse proceeds
at different rates along the three axes). Thus, as is known
\citep{Vergassola1994,Brenier2003,BernardeauVal2010b}, the onset of shell crossing is
associated with the change in sign of the Hessian determinant of $\varphi_L$ and with
the loss of convexity of the Lagrangian potential $\varphi_L$ (the Hessian
matrix is no longer positive-definite).

A well-studied dynamics that agrees with the Zeldovich dynamics
until shell crossing is provided by the ``adhesion model''
\citep{Gurbatov1989,Gurbatov1991,Vergassola1994}.
More precisely, the ``geometrical adhesion model''
\citep{BernardeauVal2010b,Valageas2010b} leads to replacing the
linear Lagrangian potential $\varphi_L(\vq)$, which defines the mapping 
$\vq\mapsto\vx$ through (\ref{xq}), by a nonlinear Lagrangian potential $\varphi$
given by the convex hull of $\varphi_L$, that is, $\varphi={\rm conv}(\varphi_L)$.
Then, particles no longer cross each other but form shocks. Moreover, this second
dynamics still coincides with the Zeldovich dynamics outside of shocks, where
$\varphi_L$ coincides with its convex hull.

In this article, we do not compute the density power spectrum associated with
this ``adhesion model'', or another explicit dynamical system, which is a difficult task.
Since we are merely interested in the amplitude of the effects associated with shell 
crossing, we simply make use of the property that shell crossing is associated with 
the loss of convexity of the Lagrangian potential $\varphi_L(\vq)$. 
Then, we note that as long as $\varphi_L(\vq)$ is strictly convex
its restriction along a line that goes through two arbitrary points  $\vq_A$
and $\vq_B$ is also strictly convex, which implies (see also
\citet{Noullez1994,Brenier2003}),
\beqa
\lefteqn{ \mbox{ before shell crossing, for any } \vq_A \neq \vq_B : } \nonumber \\
&&\hspace{2cm} [\vx(\vq_B)-\vx(\vq_A)] \cdot (\vq_B-\vq_A) > 0 .
\label{xiqi}
\eeqa
Indeed, we can choose a coordinate system so that $\vq_A=0$ and $\vq_B=(q,0,0)$,
and from strict convexity we obtain $\pl^2\varphi_L/\pl q_1^2>0$, whence
$\pl x_1/\pl q_1>0$ along the first axis and $x_1(q)-x_1(0)>0$, which reads as
(\ref{xiqi}) in its general form\footnote{In 1D the property
(\ref{xiqi}) is an obvious consequence of the absence of shell crossing,
but this is not the case in higher dimensions. Indeed, the result (\ref{xiqi}) makes use
of the constraints associated with the fact that the Lagrangian mapping
derives from a Lagrangian potential as in (\ref{xq}). This relies on
the Zeldovich dynamics and on the linear growing mode of the velocity
and displacement fields being curlfree. For the gravitational dynamics, there is no such
clear signature of the absence of shell crossing, since even in regular regions the
displacement field $\vPsi(\vq,t)$ develops rotational terms at the third order of
perturbation theory \citep{Buchert1994,BernardeauVal2008}.}.
This means that the projection of the Eulerian separation vector,
$\Delta\vx=\vx_B-\vx_A$, onto the Lagrangian vector, $\Delta\vq=\vq_B-\vq_A$,
is positive, so that $\Delta\vx$ lies in the forward half-space delimited by the plane
orthogonal to $\Delta\vq$.

Going back to the general expression (\ref{Pkxq}), the density power spectrum is
fully determined by the mean $\lag e^{\ii\vk\cdot\Delta\vx}\rag_{\vq}$,
for any $\vq$, where we note $\Delta\vx=\vx(\vq)-\vx(0)$.
To compute this quantity we can take $\vq$ along the first axis, that is, $\vq=q \,\ve_1$
with $q=|\vq|$. Then, before shell crossing we have from the constraint (\ref{xiqi}) the
property $\Delta x_1> 0$.
Therefore, we consider the ``sticky model'' defined by:
\beqa
\mbox{``sticky model''}, &  &   \mbox{for} \; \vq=|\vq| \, \ve_1 :\;\; 
\Delta x_1= \max(\Delta x_{L1},0) , \nonumber \\
&& \Delta x_2= \Delta x_{L2} , \;\; \Delta x_3= \Delta x_{L3} ,
\label{Deltax-def}
\eeqa
where $\Delta \vx_L$ is the linear Eulerian separation, as given by Eq.(\ref{Zeldef}).
Thus, this second model only differs from the Zeldovich dynamics when the
parallel linear Eulerian separation, $\Delta x_{L1}$, is negative, in which case we set
it equal to zero. This is thus a simplified version of the ``adhesion model'', as
once $\Delta x_{L1}$ reaches zero, it remains equal to zero forever.
However, the model (\ref{Deltax-def}) cannot be explicitly derived from the
``adhesion model'', since we take neither transverse directions nor larger scales
into account. Therefore, we use the more generic name ``sticky model'', to refer to this
sticking along one direction for the pair separation.

It is clear that the condition $\Delta x_{L1}<0$, where the two models
differ, is only a sufficient condition for shell crossing, but it is not necessary.
Thus, it is a local condition that does not take the ``cloud-in-cloud'' problem
into account : even though no shell crossing seems to have appeared on scale $q$ yet, it may
happen that this region is enclosed within a larger domain of size $L$ that
has already collapsed, so that particles in the smaller domain have actually 
experienced shell crossing \citep{Bond1991}.
In terms of the Lagrangian potential $\varphi_L$, which
defines the Lagrangian mapping, $\vq\mapsto\vx$, through Eq.(\ref{xq}), the absence
of shell crossing on a small domain of size $q$ means that $\varphi_L$ is equal to its
convex hull in this domain \citep{Vergassola1994,Bec2007,BernardeauVal2010b}.
However, the construction of the convex hull is a global problem, as one must
consider the behavior of $\varphi_L(\vq)$ over all the space, thereby taking into account
the ``cloud-in-cloud'' problem, while in the definition of the model (\ref{Deltax-def})
we only check a weaker condition, since we only consider the two points
$0$ and $\vq$. This means that we somewhat underestimate the effects of shell
crossing, but we can expect to obtain a reasonable estimate of their amplitude
because the probability of collapse decreases on larger scales and we perform a statistical
integration over the angles of $\vq$ in Eq.(\ref{Pkxq}).

From the previous discussions, the ``sticky model'' (\ref{Deltax-def}) and the
Zeldovich dynamics (\ref{Zeldef}) coincide before shell crossing, since then we have
$\Delta x_{L1}>0$. This implies that both theories coincide at all orders of the perturbation
theory; that is, they show the same expansions (\ref{Pnstd}) and (\ref{Pn-sigv})
over powers of $P_L$. However, they differ through nonperturbative terms, which arise
from their different behaviors after shell crossing.

From Eq.(\ref{Deltax-def}) we obtain $\Delta\Psi_1=\max(\Delta\Psi_{L1},-q)$,
for $\vq=q \, \ve_1$, and
\beqa
\lefteqn{ \lag e^{\ii\vk\cdot\Delta\vx}\rag_{q\,\ve_1} = e^{\ii\vk\cdot\vq} \,
e^{-\int \dd\vw \, [1-\cos(\vw\cdot\vq)] \, \frac{(\vk\cdot\vw)^2}{w^4} \, P_L(w)} }
\nonumber \\
&& \hspace{0.8cm} \times \, e^{\frac{1}{2} k_1^2 \sigma_{\parallel}^2}
\int_{-\infty}^{\infty}  \frac{\dd\Delta\Psi_{L1}}{\sqrt{2\pi}\sigma_{\parallel}} \, 
e^{-(\Delta\Psi_{L1})^2/(2\sigma_{\parallel}^2)} \, e^{\ii k_1 \Delta\Psi_1} ,
\label{mean-kx}
\eeqa
where we factorized the result associated with the usual Zeldovich dynamics (\ref{Pksq})
in the first two terms and we introduced the variance of the linear
longitudinal displacement,
\beqa
\sigma_{\parallel}^2(q) \! & = & \! \lag (\Delta\Psi_{L1})^2\rag 
\! = \! 2 \int\dd\vw \, [1-\cos(w_1 q)] \, \frac{w_1^2}{w^4} \, P_L(w) \\
& = & 2\sigma_v^2 -2 I_0(q) + 4 I_2(q) .
\label{sigma-par}
\eeqa
Separating the contribution from $\Delta\Psi_{L1}<-q$, we obtain for the last
two terms of Eq.(\ref{mean-kx}),
\beqa
e^{\frac{1}{2} k_1^2 \sigma_{\parallel}^2} \lag  e^{\ii k_1\Delta\Psi_1}\rag
& =& 1+ e^{\frac{1}{2} k_1^2 \sigma_{\parallel}^2} \int_{-\infty}^{-q}
\frac{\dd\Delta\Psi_{L1}}{\sqrt{2\pi}\sigma_{\parallel}} \, 
e^{-(\Delta\Psi_{L1})^2/(2\sigma_{\parallel}^2)} \, \nonumber \\
&& \hspace{-2.3cm} \times \left( e^{-\ii k_1 q}  - e^{\ii k_1 \Delta\Psi_{L1}} \right)  \\
&& \hspace{-2.7cm} = 1 \! + \frac{1}{2} \,
e^{\frac{1}{2} k_1^2 \sigma_{\parallel}^2 - \ii k_1 q}
\, \erfc\!\left(\!\frac{q}{\sqrt{2}\sigma_{\parallel}}\!\right)  - \frac{1}{2} \,
\erfc\!\left(\!\frac{q\!+\!\ii k_1 \sigma_{\parallel}^2}{\sqrt{2}\sigma_{\parallel}}\!\right) ,
\label{mean-k1x1}
\eeqa
where $\erfc(z)$ is the complementary error function (extended to the complex
plane),
\beq
\erfc(z) = \frac{2}{\sqrt{\pi}} \int_z^{\infty} \dd t \, e^{-t^2} .
\label{erfc-def}
\eeq
Then, substituting into Eqs.(\ref{mean-kx}) and (\ref{Pkxq}), we can see that the
density power spectrum of the ``sticky model'', $\Psticky(k)$, is equal to the
usual Zeldovich power spectrum obtained in Sect.~\ref{Non-linear-Zeldovich}
plus a correction term $\Psc(k)$,
\beq
\Psticky(k) = \PZel(k) + \Psc(k) ,
\label{Pnew-def}
\eeq
with
\beqa
\Psc(k) & = & \frac{1}{2} \int \frac{\dd\vq}{(2\pi)^3} \,
e^{-k^2(1-\mu^2) [\sigma_v^2-I_0(q)-I_2(q)]} \, e^{-q^2/(2\sigma_{\parallel}^2)}
\nonumber \\
&& \times \left\{ w\left(\frac{\ii q}{\sqrt{2}\sigma_{\parallel}}\right) -
w\left(\frac{\ii q-k\mu\sigma_{\parallel}^2}{\sqrt{2}\sigma_{\parallel}}\right) \right\} .
\label{Psc-def}
\eeqa
Here we introduced Faddeeva's function,
\beq
w(z) = e^{-z^2} \, \erfc(-\ii z) ,
\label{w-def}
\eeq
which satisfies the asymptotic expansion \citep{Abramowitz}
\beq
 |\arg(z)|<\frac{3\pi}{4}, \, z \rightarrow \infty : \;\; w(\ii z) \sim \frac{1}{\sqrt{\pi} z}
\left(1- \frac{1}{2z^2} + .. \right) .
\label{wz-inf}
\eeq
Expression (\ref{Psc-def}) also reads as
\beqa
\Psc(k) \!\! & = & \!\!\! \int_0^{\infty} \frac{\dd q \, q^2}{(2\pi)^2} \,
e^{-q^2/(2\sigma_{\parallel}^2)}  \int_0^1 \dd\mu \,
e^{-k^2(1-\mu^2) [\sigma_v^2-I_0-I_2]}
\nonumber \\
&& \times \Real \left\{ w\left(\frac{\ii q}{\sqrt{2}\sigma_{\parallel}}\right) -
w\left(\frac{\ii q-k\mu\sigma_{\parallel}^2}{\sqrt{2}\sigma_{\parallel}}\right) \right\} .
\label{Psc-def1}
\eeqa
As expected, we can check on Eqs.(\ref{Psc-def}) and (\ref{Psc-def1}), using the
behavior (\ref{wz-inf}), that the correction
due to shell-crossing effects is nonperturbative, in the sense that, because of the term
$e^{-q^2/(2\sigma_{\parallel}^2)}$, its expansion over powers of the amplitude of
the linear power spectrum is identically zero\footnote{This is not a large-scale or
early-time expansion, which would require knowledge of the shape of $P_L(k)$.
As for the standard perturbative expansions of Sects.~\ref{Standard-perturbative}
and \ref{Renormalized-perturbative}, one introduces for instance an auxiliary
parameter $\lambda$, such as $\delta_L \rightarrow \lambda \delta_L$ or
$P_L \rightarrow \lambda^2 P_L$, and looks for an expansion over powers
of  $\lambda$. Since $\sigma_{\parallel}^2 \propto q^2$ for $q\rightarrow 0$ if
$P_L(k)$ decreases faster than $k^{-3}$ at high $k$, as for CDM power spectra,
the integration over $q$ does not transform the exponential factor
$e^{-q^2/(2\sigma_{\parallel}^2)}$ into anomalous power laws, and the expansion
of $\Psc(k)$ over powers of the amplitude of the linear power spectrum is indeed
identically zero, hence nonperturbative.}.

At low $k$ the shell-crossing contribution (\ref{Psc-def1}) behaves as
$\Psc(k) \propto k^2$. This agrees with the generic behavior\footnote{As described in
\citet{Peebles1974}, considering momentum conservation, in addition to
matter conservation, leads to a $k^4$ tail at low $k$. However, since the ``sticky
model'' (\ref{Deltax-def}) does not satisfy momentum conservation but only matter
conservation, it gives rise to the $k^2$ tail.} associated with
small-scale redistributions of matter \citep{Peebles1974}. Thus, the ``sticky model''
provides an explicit example of a nonperturbative power spectrum, which satisfies
these generic behaviors.

\section{Numerical results in real space}
\label{Numerical}

We now describe the numerical results we obtain for the Zeldovich and
``sticky model'' power spectra, Eqs.(\ref{PkIn}) and (\ref{Pnew-def}), as well as
their common perturbative expansions.
We consider a $\Lambda$CDM cosmology, with $\Om=0.279$,
$\Omega_{\Lambda}=0.721$, $\sigma_8=0.817$, and $h=0.701$.

\subsection{Logarithmic power}
\label{Logarithmic-power}

\begin{figure*}
\begin{center}
\epsfxsize=9 cm \epsfysize=7 cm {\epsfbox{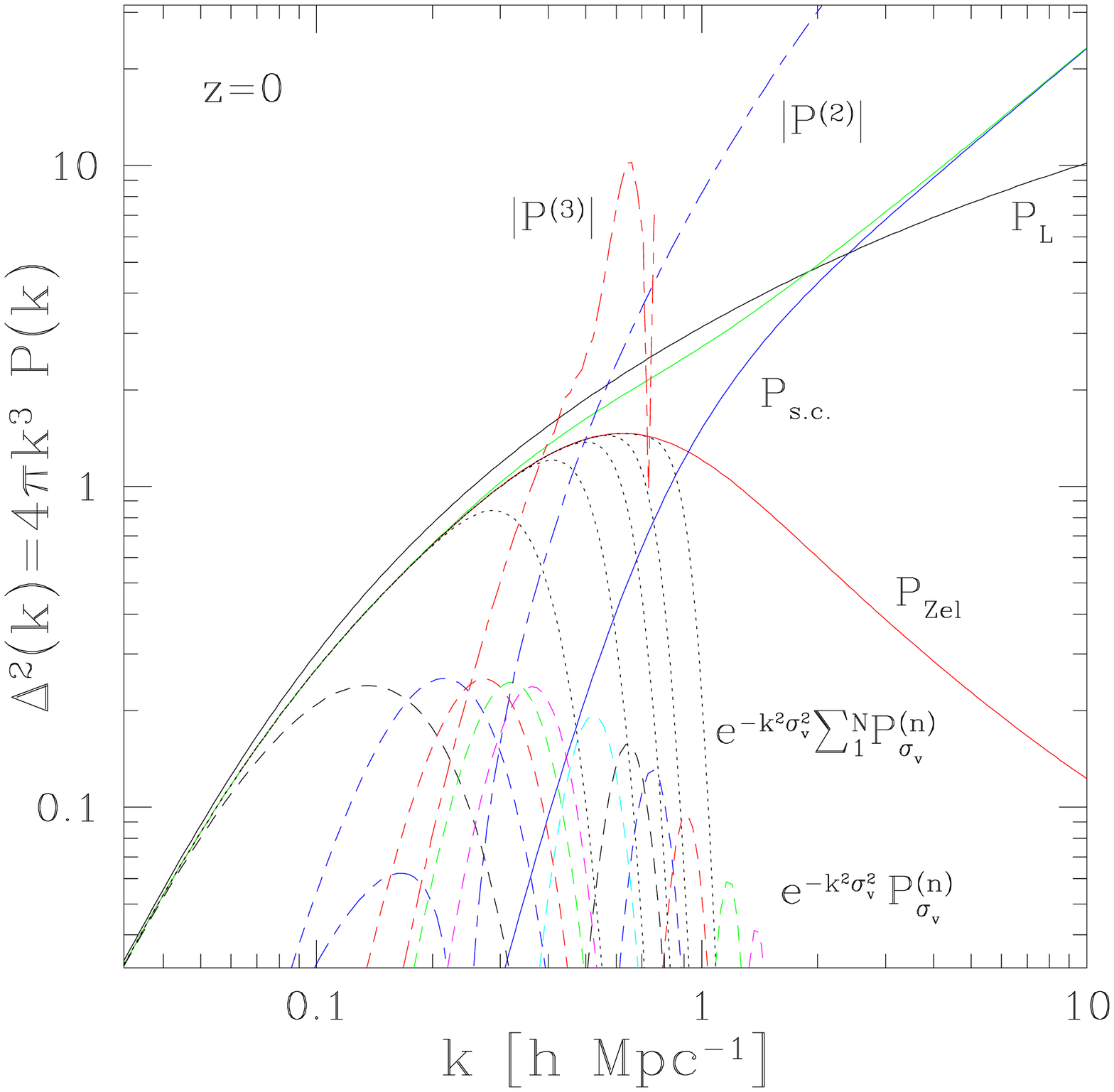}}
\epsfxsize=9 cm \epsfysize=7 cm {\epsfbox{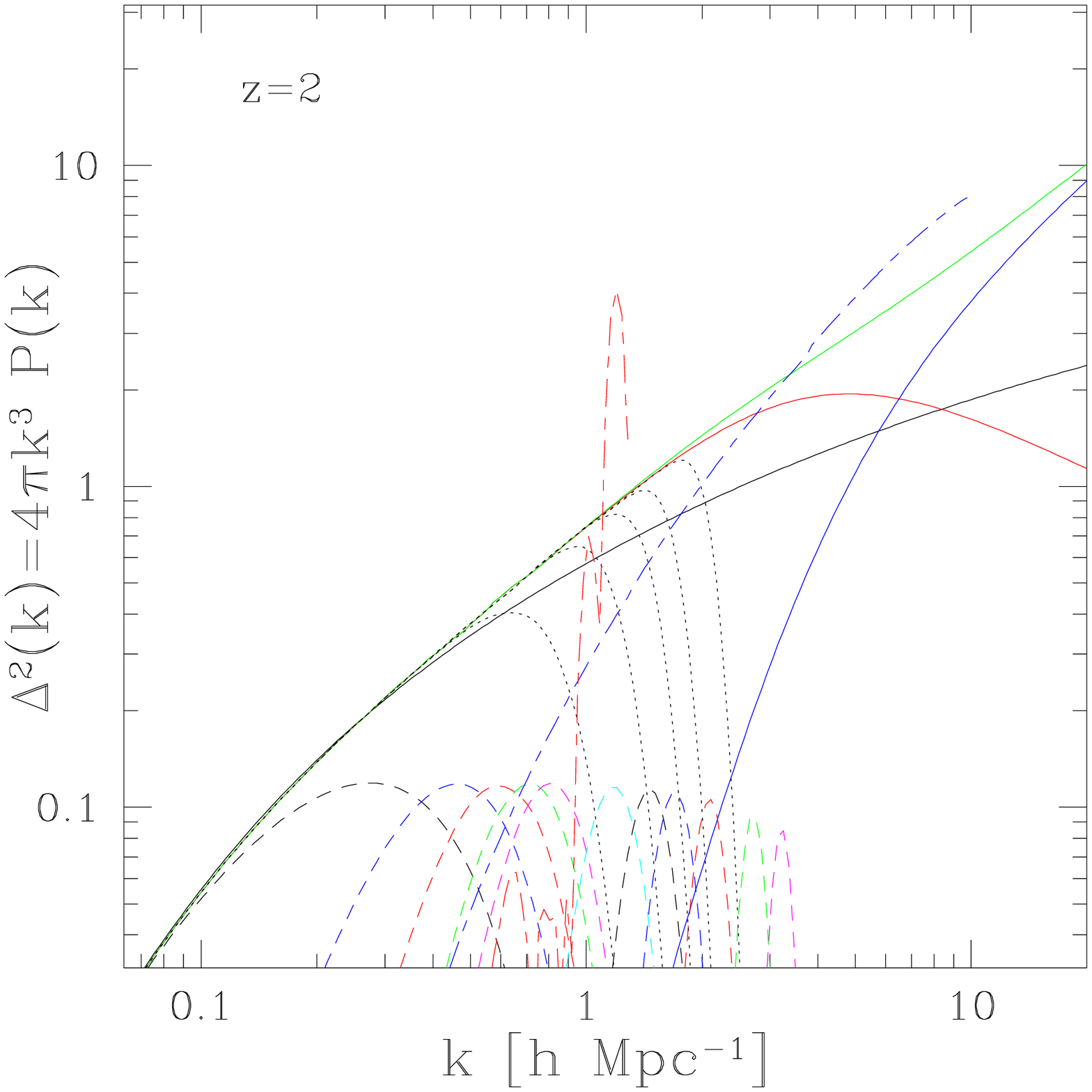}}
\end{center}
\caption{The power per logarithmic interval of $k$, as defined in Eq.(\ref{Delta2def}),
at redshifts $z=0$ (left panel) and $z=2$ (right panel). The solid lines are the linear
power spectrum ``$P_L$'', the nonlinear Zeldovich power spectrum ``$P_{\rm Zel}$'',
given by Eq.(\ref{PkIn}), the nonperturbative correction ``$\Psc$'', given by
Eq.(\ref{Psc-def1}), and the sum $P_{\rm Zel}+\Psc$ associated with the ``sticky model'',
Eq.(\ref{Pnew-def}).
The dot-dashed lines, which grow very fast and are only partly drawn, are the absolute
values of the standard
perturbative terms $|P^{(2)}|$ and $|P^{(3)}|$, from Eq.(\ref{Pnstd}).
The lower dashed lines are the ``renormalized'' perturbative terms
$e^{-k^2\sigma_v^2} P_{\sigma_v}^{(n)}$ of the expansion (\ref{Psigv}), from\
Eq.(\ref{Pn-sigv}), for $n=1$ to $5$ and $n=10, 15, 20, 30, 50$,
and $70$. The peak moves to higher $k$ as the order $n$ increases.
The dotted lines that follow the nonlinear Zeldovich power spectrum until a Gaussian
decay are the partial sums $e^{-k^2\sigma_v^2} \sum_{n=1}^{N} P_{\sigma_v}^{(n)}(k)$,
with $N=5, 10, 15, 20$, and $30$.
All terms are multiplied by the factor $4\pi k^3$ of Eq.(\ref{Delta2def}).}
\label{fig_lDk}
\end{figure*}

We show in Fig.~\ref{fig_lDk} the power per logarithmic interval of $k$, defined as
\beq
\Delta^2(k) = 4\pi k^3 P(k) ,
\label{Delta2def}
\eeq
for redshifts $z=0$ and $z=2$.
We compare the Zeldovich nonlinear power spectrum (\ref{PkIn}) with its perturbative
expansions (\ref{Pstd}) and (\ref{Psigv}) and the nonperturbative correction
(\ref{Psc-def1}). As is well known, higher order terms of the standard perturbative
expansion (\ref{Pstd}) grow increasingly fast at high $k$ with changes in sign
and large cancellations between different orders. The ``renormalized'' perturbative
expansion (\ref{Psigv}) gives positive terms (see Eq.(\ref{Pn-sigvFn})) that peak on a
well-defined range \citep{Crocce2006a} and are much easier to distinguish.
Thus, while we only plot the first three orders of the standard expansion,
$P^{(1)}=P_L$ and the absolute values $|P^{(2)}|$ and $|P^{(3)}|$, we plot the first
five orders of the ``renormalized'' expansion, $P_{\sigma_v}^{(1)}$ to
$P_{\sigma_v}^{(5)}$, as well as $P_{\sigma_v}^{(10)}$, $P_{\sigma_v}^{(15)}$,
$P_{\sigma_v}^{(20)}$, $P_{\sigma_v}^{(30)}$, $P_{\sigma_v}^{(50)}$
and $P_{\sigma_v}^{(70)}$ (always multiplied by the factor $4\pi k^3$ of
Eq.(\ref{Delta2def})).
As we go to higher orders, contributions become narrower and 
more densely packed, which implies that as we go deeper in the nonlinear regime,
we need increasingly more perturbative terms per logarithmic interval
of wavenumber.

We also plot a few partial sums of expansion (\ref{Psigv}), that is,
$e^{-k^2\sigma_v^2} \sum_{n=1}^{N} P_{\sigma_v}^{(n)}(k)$,
with $N=5, 10, 15, 20$, and $30$.
We can check that they agree with the full nonlinear power spectrum
(\ref{PkIn}) until the wavenumber associated with the peak of 
$e^{-k^2\sigma_v^2} P_{\sigma_v}^{(N)}$, after which they follow the Gaussian
decay associated with the prefactor $e^{-k^2\sigma_v^2}$.
These partial sums are also slightly more efficient than those obtained  from the
standard expansion (\ref{Pstd}) at the same order (not shown in the figure).

As expected, the nonperturbative correction (\ref{Psc-def1}) is very small
on quasi-linear scales, so that there is indeed a range where higher order terms
of the perturbative expansions (\ref{Pstd}) and (\ref{Psigv}) (i.e. beyond
the first order associated with the linear regime) are relevant.
At higher $k$, the nonperturbative correction becomes dominant and the
perturbative expansions become irrelevant, since one can no longer neglect the
physics beyond shell crossing.

In the highly nonlinear regime, we recover the well-known decay of the
Zeldovich logarithmic nonlinear power spectrum 
\citep{Schneider1995,Taylor1996,Valageas2007b}.
This is because particles escape to infinity after shell crossing
(i.e. keep moving on their straight trajectories), and these random trajectories
(in the sense of random initial conditions) erase small-scale features in the
density field. This is expressed by the Gaussian decaying factor
$e^{-k^2[\sigma_v^2-I_0-(1-3\mu^2)I_2]}$ in Eq.(\ref{PkIn}), where
the quantity within brackets is always positive as can be seen from
expression (\ref{Pkw}).
Even though this leads to a decay at high $k$, the falloff is not Gaussian
because of the integration over the Lagrangian distance $q$ \citep{Taylor1996}.
In particular, for power-law initial power spectra,
$P_L(k) \propto k^n$ with $-3<n<-1$, one finds that
$\Delta^2(k) \propto k^{3(n+3)/(n+1)}$ \citep{Valageas2007b} (the Zeldovich
dynamics is not well-defined for $n\geq -1$ because of ultraviolet divergences).
Contrary to the gravitational case, the nonlinear power spectrum
decreases faster at high $k$ for higher values of $n$. This is due to the greater
smearing out of small-scale features by the larger amplitude of the random
linear displacements at small wavelengths.

It is interesting to note that, thanks to its shell-crossing correction (\ref{Psc-def1}),
the nonlinear power spectrum (\ref{Pnew-def}) of the ``sticky model'' 
does not show this fast decay, and its logarithmic power still
increases in the nonlinear regime. This is due to the prescription (\ref{Deltax-def}),
which in a sense prevents particles from escaping to infinity in one direction,
as they stick together. This approximately models the formation
of Zeldovich pancakes, associated with collapse along one axis (although there is
no precise relationship, since Eq.(\ref{Deltax-def}) is only a statistical model
and does not consider the ``cloud-in-cloud'' problem).
Thus, small-scale structures are no longer erased, since we keep a trace of planar
features. This is expressed by the factor $(1-\mu^2)$ in the exponential argument
in expression (\ref{Psc-def1}), which suppresses the Gaussian decaying term of the
form $e^{-k^2}$ for $\mu \simeq 1$. This gives a width $\Delta \mu \sim k^{-2}$,
hence $\Psc(k) \sim k^{-2}$ at high $k$, as would be the case for a density
field where planar objects are the relevant nonlinear structures (i.e. bi-dimensional
structures, as opposed to pointlike masses or lines for instance).
Contrary to the Zeldovich power spectrum
there is no dependence on $n$ for the high-$k$ slope, for power-law linear 
power spectra $P_L(k) \propto k^n$ with $-3<n<-1$.
This gives the universal asymptote $\Delta^2_{\rm sticky}(k) \propto k$
for the nonlinear logarithmic power of the ``sticky model'' at high $k$.

\subsection{Perturbative and nonperturbative contributions}
\label{corrections}

\begin{figure*}
\begin{center}
\epsfxsize=8.5 cm \epsfysize=6 cm {\epsfbox{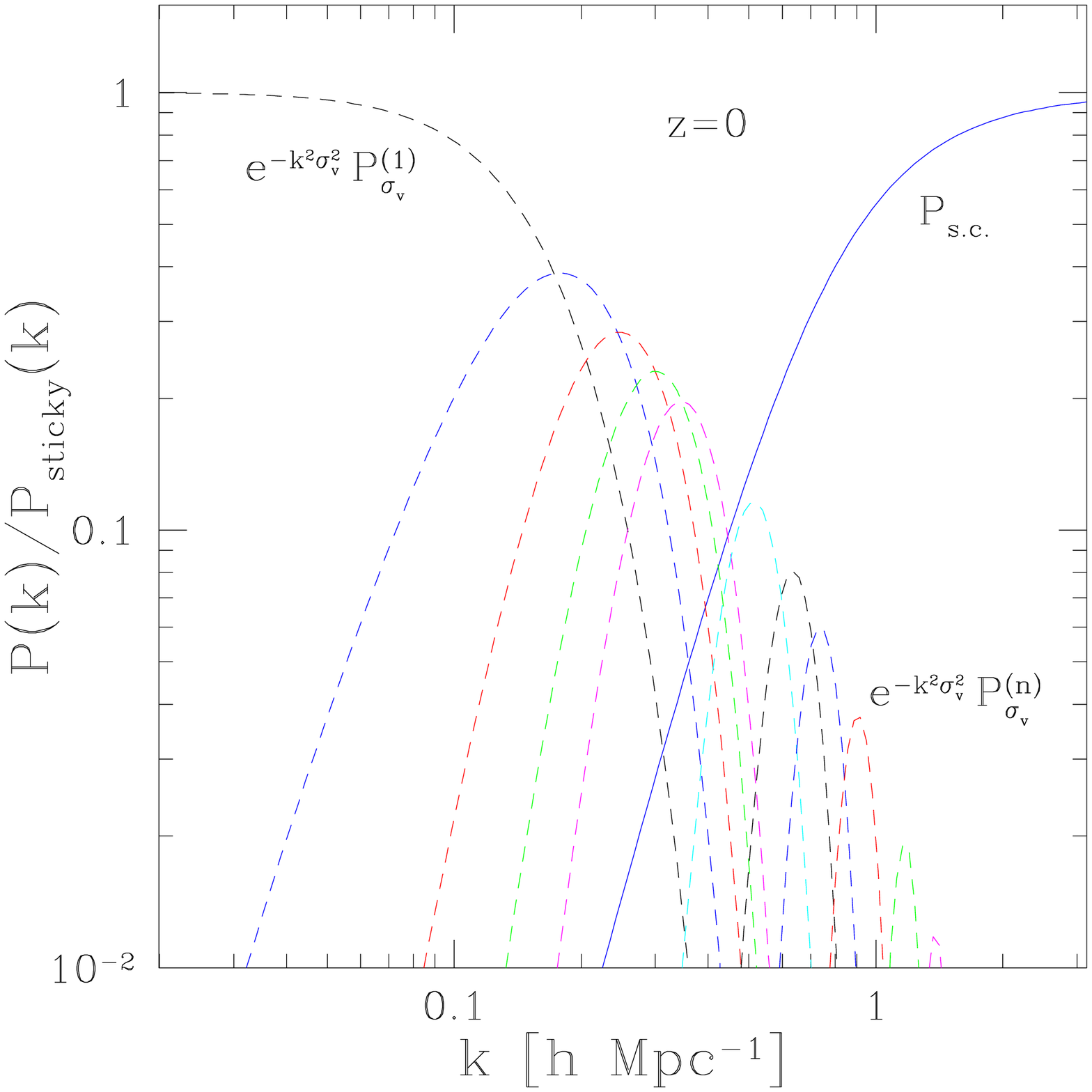}}
\epsfxsize=8.5 cm \epsfysize=6 cm {\epsfbox{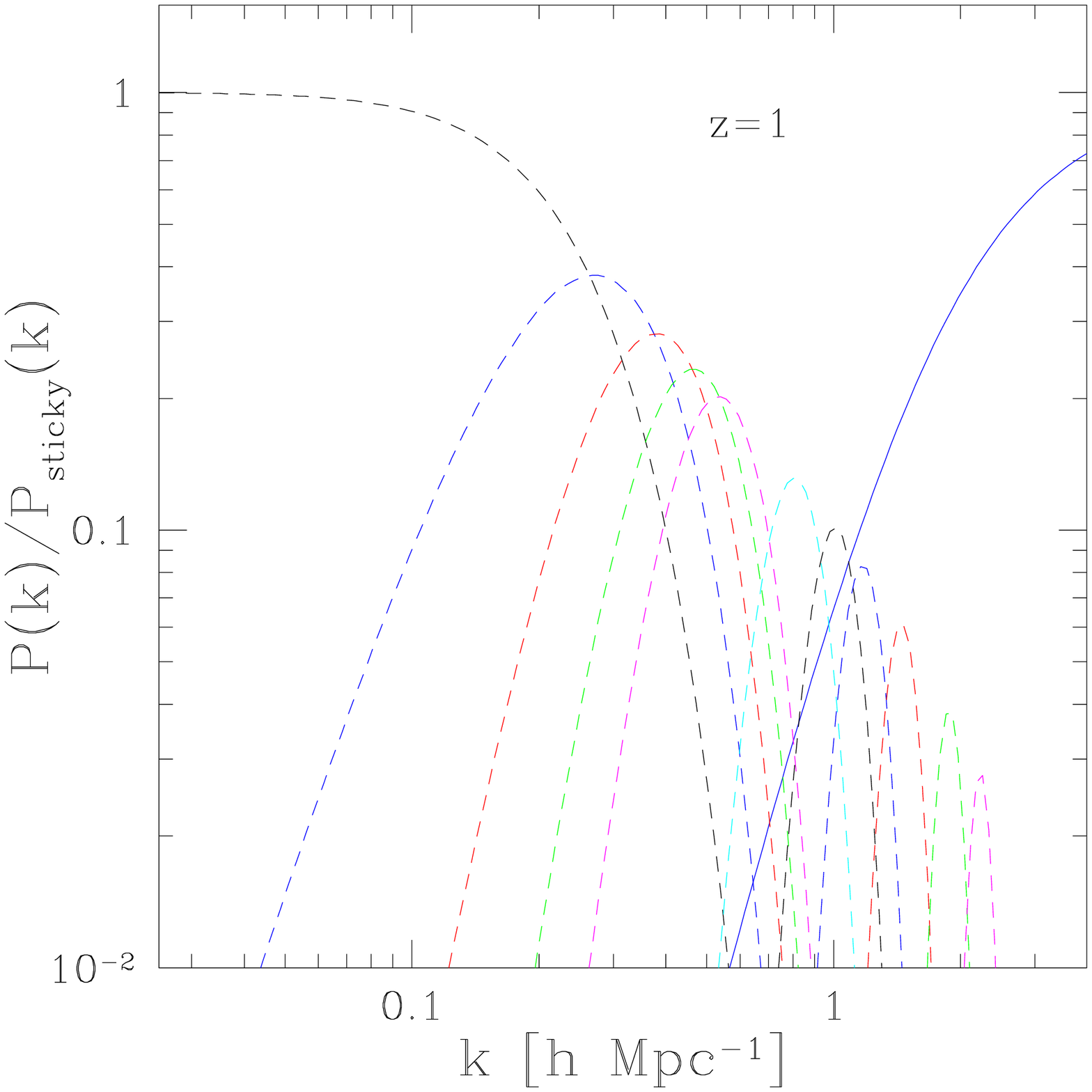}}\\
\epsfxsize=8.5 cm \epsfysize=6 cm {\epsfbox{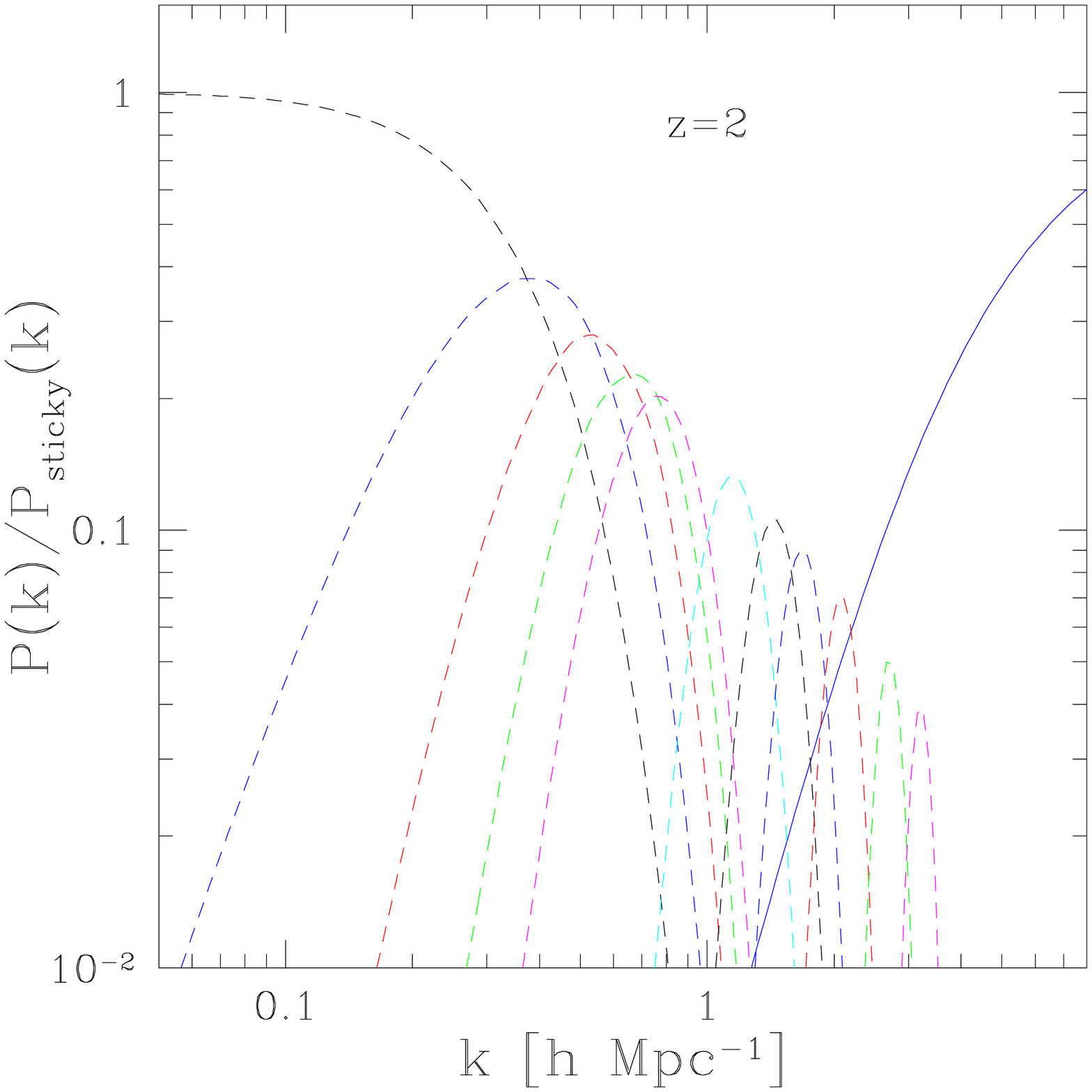}}
\epsfxsize=8.5 cm \epsfysize=6 cm {\epsfbox{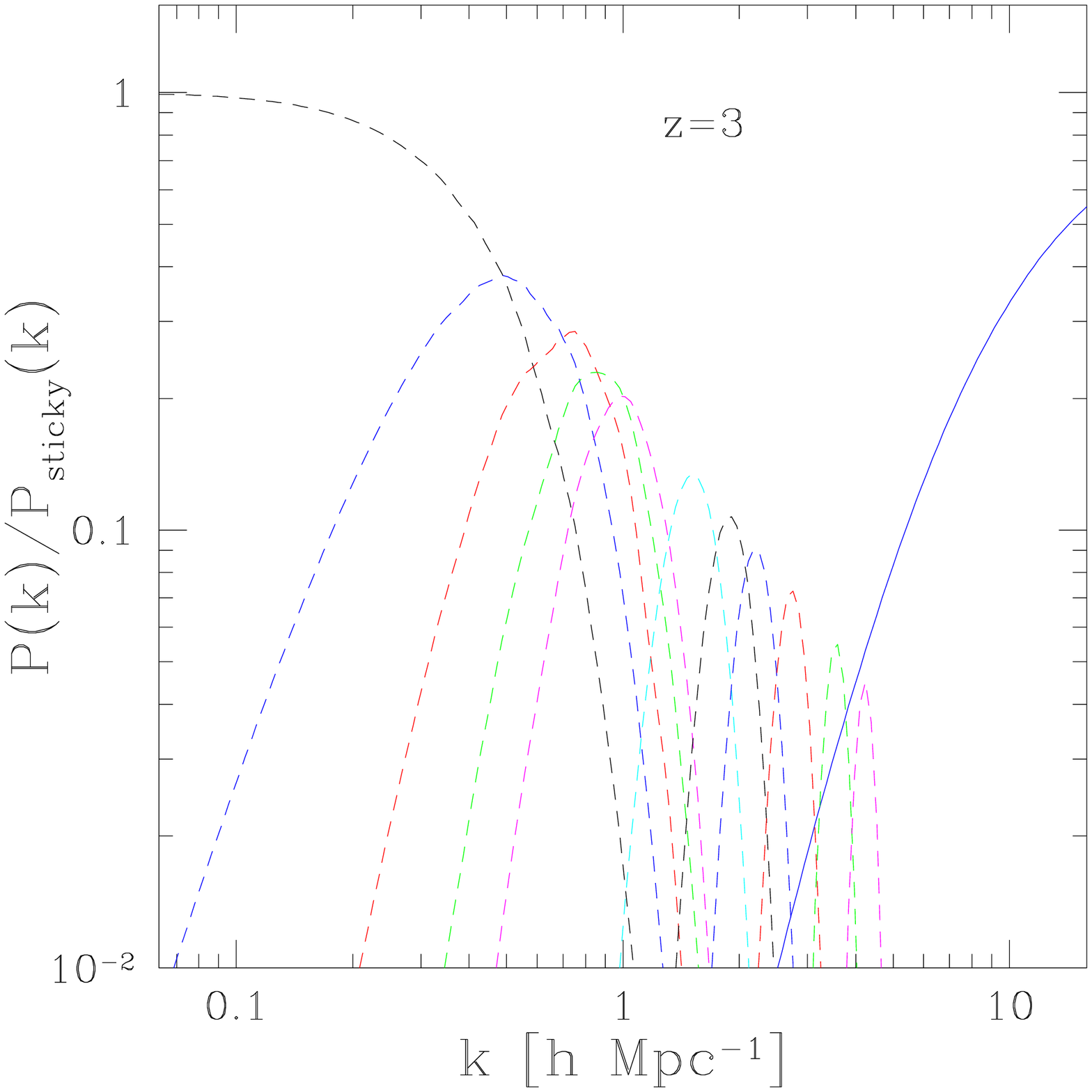}}
\end{center}
\caption{The dashed lines are the ratios of the ``renormalized'' perturbative terms
of the expansion (\ref{Psigv}) to the nonlinear ``sticky model'' power spectrum,
$e^{-k^2\sigma_v^2} P_{\sigma_v}^{(n)}(k)/P_{\rm sticky}(k)$, for $n=1$ to $5$ and
$n=10, 15, 20, 30, 50$, and $70$.
The solid line is the ratio of the nonperturbative correction to the nonlinear power
spectrum, $\Psc(k)/P_{\rm sticky}(k)$. The four panels correspond to redshifts
$z=0,1,2$, and $3$ (with a shift of the plot towards higher $k$ at higher $z$).}
\label{fig_lrDk}
\end{figure*}

To clearly see the range of scales where perturbative schemes
are relevant, we plot in Fig.~\ref{fig_lrDk} the ratios of the successive terms of
the ``renormalized'' perturbative expansion (\ref{Psigv}) with respect to the
nonlinear power spectrum of the ``sticky model'',
$e^{-k^2\sigma_v^2} P_{\sigma_v}^{(n)}(k)/P_{\rm sticky}(k)$, for
$n=1$ to $5$, as well as $n=10, 15, 20, 30, 50$, and $70$.
We also plot the ratio $\Psc(k)/P_{\rm sticky}(k)$, to compare with the amplitude of
the nonperturbative correction associated with shell-crossing effects.
We consider the four redshifts $z=0,1,2$, and $3$.
Here we focus on the ``renormalized'' perturbative expansion (\ref{Psigv}), rather than
on the standard expansion (\ref{Pstd}), to avoid the interferences brought by the
changes in sign and the cancellations between various terms.

We can see that the potential of perturbative expansions grows at higher redshift
for a $\Lambda$CDM power spectrum, because of the change in
slope of $P_L(k)$. Thus, at $z=3$ we can go up to the order $n=66$ 
before the perturbative term becomes subdominant with respect to the nonperturbative
correction, while at $z=0$ the crossover takes place at $n=9$, as the term
$e^{-k^2\sigma_v^2} P_{\sigma_v}^{(10)}(k)$ is the lowest order one that is fully
below the nonperturbative correction $\Psc(k)$.
This agrees with the observation that in the gravitational
case perturbative schemes (and resummation approaches) seem to
fare better at higher $z$ \citep{Carlson2009}.
We can see that perturbative schemes are relevant over roughly one decade over
$k$. Thus, if we require an accuracy of $1\%$ at
$z=0$, linear theory is sufficient up to $k_{1\rm loop} \sim 0.033 h$ Mpc$^{-1}$, while
higher order perturbative terms allow reaching $k_{\rm s.c.} \sim 0.23 h$ Mpc$^{-1}$.
At higher $k$ one must take the nonperturbative correction associated
with shell-crossing effects into account, which implies going beyond the fluid approximation
and requires new approaches. 

To help the reader, we give in Table~\ref{Table_range}
the wavenumber $k_{1\rm loop}$, below which the linear term is enough
to reach a $1\%$ or $10\%$ accuracy, for the four redshifts shown in Fig.~\ref{fig_lrDk}.
We also give the wavenumber $k_{\rm s.c.}$ beyond which the nonperturbative
correction is required to reach an accuracy of $1\%, 10\%$, or $50\%$ (in units of
$P_{\rm sticky}(k)$). Thus, the interval $[k_{1\rm loop},k_{\rm s.c.}]$ gives the range
where perturbation theories based on the fluid description are relevant.
Of course, this range shifts to higher $k$ at higher redshift. It is interesting to
note that this range is also broader at higher redshift, as the slope of the CDM linear power
spectrum on the relevant scales changes slowly.
The last column gives the last order, $n_{\rm s.c.}$, of the ``renormalized''
perturbative expansion that is not fully below the nonperturbative term. As noticed
above in Fig.~\ref{fig_lrDk}, this expansion order is significantly higher at higher redshift.
This corresponds to a greater potential for perturbative schemes. However,
$n_{\rm s.c.}$ grows faster than the logarithmic width of the perturbative range,
$[\ln k_{1\rm loop},\ln k_{\rm s.c.}]$. Indeed, as shown in Fig.~\ref{fig_lrDk},
peaks associated with higher order perturbative terms are increasingly narrow
on the $\ln k$ axis. This implies that to multiply the upper wavenumber
$k$, defined by a fixed accuracy, by a given amount, one needs to add an increasingly
greater number of new perturbative terms.

\begin{table}
\begin{center}
\begin{tabular}{c||c|c|c||c}
$z$ & $\%$ & $k_{1\rm loop} \; [h {\rm Mpc}^{-1}]$ & $k_{\rm s.c.} \; [h {\rm Mpc}^{-1}]$ & $n_{\rm s.c.}$  \rule[-0.3cm]{0cm}{0.6cm} \\ \hline\hline
 & $1\%$ & 0.033 & 0.23 &  \rule[-0.2cm]{0cm}{0.6cm} \\
0 & $10\%$ & 0.082 & 0.45 & 9 \rule[-0.2cm]{0cm}{0.4cm} \\ 
 & $50\%$ & & 0.9 &  \rule[-0.2cm]{0cm}{0.4cm} \\ \hline
 & $1\%$ & 0.043 & 0.44 &  \rule[-0.2cm]{0cm}{0.6cm} \\
1 & $10\%$ & 0.11 & 1.1 & 18 \rule[-0.2cm]{0cm}{0.4cm} \\
 & $50\%$ & & 2.2 &  \rule[-0.2cm]{0cm}{0.4cm} \\ \hline
 & $1\%$ & 0.057 & 1.2 & \rule[-0.2cm]{0cm}{0.6cm}  \\
2 & $10\%$ & 0.14 & 2.3 & 37 \rule[-0.2cm]{0cm}{0.4cm} \\
 & $50\%$ & & 6.4 &  \rule[-0.2cm]{0cm}{0.4cm} \\ \hline
 & $1\%$ & 0.07 & 2.2 & \rule[-0.2cm]{0cm}{0.6cm} \\
3 & $10\%$ & 0.18 & 5.2 & 66 \rule[-0.2cm]{0cm}{0.4cm} \\ 
 & $50\%$ & & 10.4 &  \rule[-0.2cm]{0cm}{0.4cm} \\
\end{tabular}
\end{center}
\caption{The dependence on redshift $z$ of the wavenumbers where the ratios
$e^{-k^2\sigma_v^2} P_{\sigma_v}^{(2)}(k)/P_{\rm sticky}(k)$ (at $k_{1\rm loop}$) and
$\Psc(k)/P_{\rm sticky}(k)$ (at $k_{\rm s.c.}$) reach $1\%$, $10\%$, and $50\%$.
The last column gives the last order $n_{\rm s.c.}$ of the ``renormalized'' perturbative
expansion that is not fully below the nonperturbative term.}
\label{Table_range}
\end{table}

In practice, we do not expect that perturbative terms will be computed up to
such high orders, since in the case of the gravitational dynamics this would involve
multidimensional integrals that are beyond the reach of current numerical
possibilities. However, resummation schemes allow one to consider
parts of such higher order terms (actually, an infinite number of diagrams that
contribute to terms of all orders). Then, the hope is that such methods
can efficiently sum most of the contributions of higher order terms
and accelerate the convergence on weakly nonlinear scales.
The comparison displayed in Fig.~\ref{fig_lrDk} shows the potential of such methods
(i.e. the best result one can obtain, for the ``sticky model'' considered here), which
appears to be quite significant at higher redshifts, $z\geq 1$.

Since the Zeldovich dynamics provides a reasonable approximation of the
gravitational dynamics down to weakly nonlinear scales \citep{Coles1993,Pauls1995},
and its accuracy can be improved by implementing the ``adhesion model''
that only differs after shell crossing 
\citep{Weinberg1990,Melott1994,Sathyaprakash1995},
we can expect that to a large extent these results
still apply to the gravitational case.
In particular, while we find that at $z=0$ the nonperturbative correction to the
density power spectrum is around $1\%$ at $k\sim 0.23 h$ Mpc$^{-1}$ and
$50\%$ at $k\sim 0.9 h$ Mpc$^{-1}$, \citet{Afshordi2007} finds the wavenumbers
$k\sim 0.1 h$ Mpc$^{-1}$ and $k\sim 0.85 h$ Mpc$^{-1}$ with a
phenomenological ``sticky halo model''. It is comforting that these two very different
approaches give similar results. Then, the property that the range where
perturbative schemes are relevant is greater at $z=3$ than at $z=0$, with a higher
order $n_{\rm s.c.}$, should remain valid.

This is confirmed by the analysis of unified models that combine perturbation
theories with halo models, for the gravitational case. Thus, as shown in
\citet{Valageas2010a}, at higher redshift the intermediate range,
where the power spectrum departs from one-loop perturbation theory but is not yet
well described by the ``one-halo'' contribution, becomes wider. There, the ``one-halo''
contribution plays the
role of the nonperturbative contribution (\ref{Psc-def1}), as it is also fully
nonperturbative and decays at low $k$ as $k^2$ (if we only consider
matter conservation, see the discussion in \citet{Valageas2010a}).
Then, the broadening of this intermediate range also suggests that higher orders
of perturbation theory become relevant, as explicitly found in Figs.~\ref{fig_lDk}
and \ref{fig_lrDk}.

In the context of cosmological reconstruction\footnote{
There, the problem is to reconstruct the past dynamical history of a given region
of the sky from the knowledge of its present density field. A key
observation is that in the linear growing mode, the velocity field is curlfree
and related to the linear gravitational potential by a relation of the form
$\vv_L \propto -\nabla\Phi_L$, while the linear density field is given by
the Poisson equation, $\delta_L \propto \Delta \Phi_L$. Then, the number of
unknowns (the linear gravitational potential) is equal to the number of constraints
(the present nonlinear density field), so that it makes sense to look for a
reconstruction defined by the nonlinear density field alone (but this does not
ensure uniqueness). The Monge-Amp\`ere-Kantorovich method 
makes use of an additional assumption, that the Lagrangian map, $\vq\mapsto\vx$,
can be derived from a convex potential $\varphi(\vq)$, 
as in Eq.(\ref{xq}), to derive a unique solution to the ``displacement reconstruction'',
i.e. the inverse map $\vx\mapsto\vq$ \citep{Brenier2003,Mohayaee2006}.
Such methods cannot describe multi-streaming, and by estimating the scale
where shell crossing effects can no longer be neglected,  we can evaluate the scale
down to which these reconstruction schemes can apply.}, a comparison
with $N$-body simulations \citep{Mohayaee2006} shows that the
Monge-Amp\`ere-Kantorovich method is able to recover the nonlinear displacement
field down to $\sim 3 h^{-1}$ Mpc at $z=0$, which corresponds roughly to
$k \sim 2 h$ Mpc$^{-1}$. We can see from Table~\ref{Table_range} that this is
a very good result, as going to smaller scales requires taking shell crossing into
account (in fact, at $k=2 h$ Mpc$^{-1}$ nonperturbative corrections have already
started to dominate).
Thus, the Monge-Amp\`ere-Kantorovich method appears to be close to
optimal at $z=0$, because it goes as far as any scheme that disregards
shell-crossing effects can be expected to go.
This can be understood partly from the fact that relatively few orders of perturbations 
theory are relevant at $z=0$ (since $n_{\rm s.c.}=9$), so that it may not help much
to explicitly include the effects of higher order terms.

Reconstruction techniques are also used to sharpen the acoustic peak of the
real-space correlation function or to restore the harmonics of the oscillations
of the power spectrum, in order to improve cosmological distance measurements
and constraints on dark energy \citep{Eisenstein2007,Seo2010}.
Then, one can read in Table~\ref{Table_range} the wavenumber $k_{\rm s.c.}$ up
to which one can hope to recover these baryon acoustic oscillations.
At $z=0$ we can see that present schemes, which are based on the linear displacement
field and manage to reach $\sim 0.2 h$ Mpc$^{-1}$ (not necessarily for the
amplitude but at least for the shape and position of the oscillations) are not far from
the upper  bound, as could be expected from only a few orders of
perturbation theory being relevant ($n_{\rm s.c.}=9$). At $z=3$ it seems that one
could go much beyond present schemes (which do not go much farther than
$0.25 h$ Mpc$^{-1}$), in agreement with $n_{\rm s.c.}=66$, which
means that higher orders of perturbation theory are relevant.
However, for the specific purpose of measuring cosmological distances from the
baryon acoustic oscillations, the potential is limited by the relative amplitude
of the oscillations of the linear power spectrum itself decreasing at higher $k$, so
that even a very good reconstruction would not greatly enhance the signal-to-noise
ratio. Nevertheless, pushing to higher orders (e.g., through resummation schemes)
remains useful for other purposes, such as weak-lensing studies.

\subsection{Rise of power at the transition}
\label{Rise}

The increase of power on the transition scale to nonlinearity shown in Fig.~\ref{fig_lDk}
(especially in the left panel at $z=0$)
is reminiscent of a similar feature observed for the gravitational dynamics
\citep{Hamilton1991,Peacock1996}.
This is usually interpreted from a Lagrangian point of view inspired by the spherical
collapse dynamics. Thus, \citet{Padmanabhan1996} argues that, on
these intermediate scales, one has $\xib(x) \propto \xib_L(q)^3$ in real space
with $q^3=x^3(1+\xib)$. More generally, in Fourier space one writes for the
nonlinear power per logarithmic interval of wavenumber, $\Delta^2(k)$, the
parametric system \citep{Peacock1996}
\beqa
k_L & = & [1+\Delta^2(k) ]^{-1/3} k , \label{kL} \\
\Delta^2(k) & = & f[ \Delta_L^2(k_L) ] , \label{f-def} 
\eeqa
with a function $f$ to be determined.
These relations express the conservation of matter,
since the Lagrangian scale $q\sim 1/k_L$ collapses down to the Eulerian scale $x\sim 1/k$.
The linear regime implies that $f(x) \simeq x$ for $x \ll 1$, whereas in the highly
nonlinear regime the stable-clustering ansatz \citep{Peebles1982} gives the
scaling $f(x) \sim x^{3/2}$ for $x \gg 1$. At the transition one observes a sharper
growth, which is consistent with $f(x) \sim x^3$ \citep{Padmanabhan1996}.
In practice, one builds a fitting formula for $f(x)$ to match numerical simulations and
to account for the dependence on the shape of the linear power spectrum.

In any case, such models usually estimate the shape of the nonlinear two-point
correlation function or of the nonlinear power spectrum by considering the collapse
of a ``typical'' overdensity \citep{Padmanabhan1996,Valageas1997} (or merely obtaining
$f(x)$ from simulations without further interpretation).
It is interesting to note that this collapse also takes place within the Zeldovich dynamics
studied here. In particular, the function $\delta=\cF(\delta_L)$
that describes the spherical collapse is no longer given by cycloids \citep{Peebles1980}
but by the simple expression $\cF(\delta_L)= 1/(1-\delta_L/3)^3-1$
\citep{Bernardeau1995,Valageas2009b}.  (Collapse to a point is delayed
from $\delta_c \simeq 1.686$ to $\delta_c =3$, since the motion does not accelerate
as the gravitational potential well becomes deeper.)

\begin{figure}
\begin{center}
\epsfxsize=8.5 cm \epsfysize=6 cm {\epsfbox{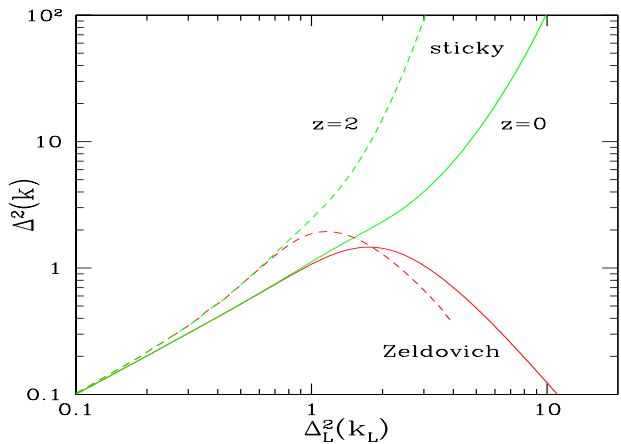}}
\end{center}
\caption{The nonlinear power per logarithmic interval of wavenumber, $\Delta^2(k)$,
as a function of the linear power $\Delta_L^2(k_L)$, at the Lagrangian wavenumber
$k_L$ defined by Eq.(\ref{kL}). The lower curves are obtained from the Zeldovich
power spectrum (\ref{PkIn}), while the upper curves correspond to the power spectrum
(\ref{Pnew-def}) of our ``sticky model''. The solid lines are for the redshift $z=0$ and
the dashed lines for $z=2$.}
\label{fig_lfDk}
\end{figure}

We show in Fig.~\ref{fig_lfDk} the functions $\Delta^2(k) = f[ \Delta_L^2(k_L) ]$
defined by the system (\ref{kL})-(\ref{f-def}) for the Zeldovich dynamics and the
``sticky model''. Indeed, from the knowledge of the nonlinear power $\Delta^2(k)$
we obtain the Lagrangian wavenumber $k_L$ from Eq.(\ref{kL}) and next
$\Delta_L^2(k_L)$. For a given Eulerian wavenumber $k$, the Lagrangian
wavenumbers $k_L$ obtained for both models, Zeldovich dynamics and
``sticky model'', are different.
Then, Fig.~\ref{fig_lfDk} shows that, within the
Zeldovich dynamics, this growth of the nonlinear density contrast through the spherical
collapse is not sufficient to build up the increase in power 
on these mildly nonlinear scales, although one can see a modest rise at $z=2$.
Indeed, the shape of the nonlinear power spectrum is quickly governed by
the decay that takes place at higher $k$, due to the escape of particles
beyond shell crossing that erases small-scale features, as recalled in
Sect.~\ref{Logarithmic-power}. In contrast, such a growth is clearly seen in the 
``sticky model''.

This means that such arguments, based on the evolution of a ``typical'' overdensity
through the spherical collapse dynamics, are not good enough to explain the shape of the
power spectrum on these scales since they clearly fail for the Zeldovich dynamics.
Thus, one needs to consider the behavior of a whole range of typical
density fluctuations (i.e., truly perform the Gaussian average over the initial conditions),
which includes a significant number of configurations where shell crossing has already
taken place. Then, one cannot neglect the dependence on their behavior after shell
crossing, and whereas the free escape associated with the Zeldovich dynamics
is sufficient to erase any growth of power on the transition scale, a trapping of particles
allows such a growth. In the ``sticky model'' studied here, this trapping after shell
crossing is a simple sticking along one direction, as defined in (\ref{Deltax-def}),
which eventually leads to a universal $k^{-2}$ tail at very high $k$ for $\Psticky(k)$, as
noticed in Sect.~\ref{Logarithmic-power}. In terms of the function
$f[ \Delta_L^2(k_L) ]$, this corresponds to a large-$x$ scaling
$f_{\rm sticky}(x) \propto x^{3/[2(n+3)]}$ from the system (\ref{kL})-(\ref{f-def}).

In the gravitational case, this regime corresponds to the virialization within 3D bound
structures, with an asymptotic tail at high $k$ that is poorly known.
In particular, although overdensities seem to form halos with an almost universal
profile \citep{NFW1997}, there is no
first-principle derivation of the high-$k$ exponent of the density power spectrum,
and its degree of universality is not well known.
However, especially in the case of CDM-like power spectra that decay as
$P_L(k)\propto k^{-3}$ at high $k$, halos are rather smooth with a low amount of
substructures. Then, the nonlinear logarithmic power spectrum does not seem to show
a strong high-$k$ tail, such as the universal forms $\Delta^2(k)\propto k^{3}$, 
$\propto k^{2}$ or $\propto k^{1}$, associated with pointlike masses, lines, or sheets,
but appears to be consistent with a logarithmic or shallow slope asymptote
\citep{Smith2003}.
This means that the steep rise in the function $\Delta^2(k) = f[ \Delta_L^2(k_L) ]$
observed in the intermediate regime, $1<\Delta^2(k)<100$, breaks down 
at $\Delta^2(k) \sim 100$ to reach this slow growth at higher $k$, in contrast to
the ``sticky model'' considered here where there is a regular transition from the
linear regime, where $f_{\rm sticky}(x) \sim x$, to the highly nonlinear regime, where
$f_{\rm sticky}(x) \propto x^{3/[2(n+3)]}$. 

On the other hand, in agreement with the fact that the linear density contrast
associated with full spherical collapse to a point is delayed from $\delta_c \simeq
1.686$ to $3$, the rise obtained in Fig.~\ref{fig_lfDk} at $\Delta^2_L(k) \sim 1$
appears at somewhat higher values of $\Delta^2_L(k)$ than would be the case
for the gravitational dynamics, as can be checked by comparison with the
results of $N$-body simulations \citep{Smith2003}. In this respect, the
spherical collapse captures some of the properties of the early rise of
$\Delta^2(k)$.

\section{Redshift-space power spectrum}
\label{Redshift-space-power-spectrum}

The density field observed from galaxy surveys is distorted by their peculiar
velocities, which introduce an error in the measure of their radial position
\citep{Jackson1972,Kaiser1987}.
Indeed, both the mean Hubble flow and the peculiar radial velocity of a galaxy
contribute to the redshift of the observed emission and absorption lines.
On the other hand, the measure of the angular positions on the sky are not
influenced by these peculiar velocities. In a ``plane-parallel'' approximation,
where the line of sight has a fixed direction $\ve_z$, radial and angular modes
correspond to longitudinal and transverse wavenumbers $\kpar$ and $\vkperp$.
Then, the power spectrum measured along the transverse direction
(i.e. $\kpar=0$) is not affected by these ``redshift distortions''
and is equal to the real-space power spectrum considered in the previous
sections, whereas the power spectrum measured along the longitudinal direction
(i.e. $\vkperp=0$) is modified. In particular, this means that
the power spectrum is no longer isotropic.

This implies that some additional information is contained in the
redshift-space power spectrum, since by comparing the longitudinal and transverse
components, we can derive some information on the velocity field and obtain
further constraints on cosmology.
For instance, within linear theory, comparing the redshift-space power spectrum
over several directions \citep{Kaiser1987} or expanding its angular
dependence over Legendre polynomials and comparing the first few multipoles
\citep{Hamilton1992}, one can constrain the ratio $f$ of the linear growth rate
to its value in a critical density universe, hence $\Om$ since $f\simeq \Om^{0.6}$.  
On the other hand, while clustering along the transverse directions of statistical
standard rulers such as BAO gives the angular distance, real-space clustering along the
longitudinal direction gives the Hubble rate, so that complementary information
can be derived from both directions, and one can use the longitudinal/transverse
ratio to perform the Alcock-Paczynski test \citep{Alcock1979}.

Thus, it is interesting to extend the analysis presented in the previous sections to
the redshift-space power spectrum. Rather than expanding on multipoles, we
focus on the clustering along the radial direction, as compared with the
transverse directions.
Of course, another effect that comes into play in galaxy surveys is the bias,
which may show some scale dependence. However, we do not study this effect
here, as this is a rather different process.

\subsection{Nonlinear Zeldovich power spectrum in redshift space}
\label{Non-linear-Zeldovich-redshift}

We first recall the nonlinear redshift-space power spectrum associated with the
Zeldovich dynamics. The redshift-space coordinate $\vs$ of a galaxy is
\beq
\vs= \vx + \frac{\ve_z\cdot\vv}{aH} \, \ve_z ,
\label{sdef}
\eeq
where $\vx$ is its comoving position, $\vv=a\dot{\vx}$ its peculiar velocity,
and $\ve_z$ the unit vector of the line of sight.
Within the Zeldovich dynamics (\ref{Zeldef}) the peculiar velocity is
\beq
\vv = a \dot{\vPsi}_L = \frac{a\dot{D}}{D} \, \vPsi_L ,
\label{vdef}
\eeq
leading to
\beq 
\vs(\vq,t) = \vq + \vPsi_L + f \, (\ve_z\cdot\vPsi_L) \, \ve_z ,
\label{sq}
\eeq
where $D$ is the linear growth factor and $f(z)= \dd\ln D/\dd\ln a$.
In the following we use a ``plane-parallel'' approximation and we denote the
longitudinal and transverse directions to the line of sight by the
subscripts $\parallel$ and $\perp$. Thus we have
\beq
s_{\parallel} = q_{\parallel} + (1+f) \Psi_{L\parallel} , \;\;\;
\vs_{\perp} = \vq_{\perp}+ \vPsi_{L\perp} .
\label{sd-perp}
\eeq
The conservation of matter reads again as $\rho^{s}(\vs)\dd\vs=\rhob\dd\vq$,
where we denote the redshift-space quantities by a superscript ``s'', and as in
Eqs.(\ref{rhox})-(\ref{Pkxq}) the redshift-space power spectrum reads as
\beq
P^s(\vk) = \int\frac{\dd\vq}{(2\pi)^3} \, \lag e^{\ii \vk \cdot [\vs(\vq)-\vs(0)]} \rag .
\label{Pkq-s}
\eeq
Following \citet{Taylor1996}, it is convenient to introduce the vector $\vK$,
which is the wavevector $\vk$ stretched by $(1+f)$ along the line of sight,
\beq
\vK =  (1+f) k_{\parallel} \, \ve_z + \vkperp  .
\label{Kdef}
\eeq
Then Eq.(\ref{Pkq-s}) writes as (see Eq.(\ref{PkIn}))
\beqa
P^s(\vk) & = & \int\frac{\dd\vq}{(2\pi)^3} \, e^{\ii\vk\cdot\vq} \, 
\lag e^{\ii\vK\cdot[\vPsi_L(\vq)-\vPsi_L(0)]} \rag \nonumber \\
& = & \int\frac{\dd\vq}{(2\pi)^3} \, e^{\ii\vk\cdot\vq} \,
e^{-K^2[\sigma_v^2-I_0(q)-(1-3\mu_{Kq}^2)I_2(q)]} ,
\label{Psk-Iq}
\eeqa
where $\mu_{Kq}=(\vK\cdot\vq)/(Kq)$. Using spherical coordinates about the
vector $\vK$, expanding parts of the exponentials and using Eq.(\ref{int-mu})
we obtain
\beqa
P^s(\vk) \! & = & \! \int_0^{\infty}\frac{\dd q \, q^2}{2\pi^2} \,
e^{-K^2[\sigma_v^2-I_0(q)+2I_2(q)]} \sum_{\ell,m=0}^{\infty}
\frac{(\ell+m)!}{\ell ! \, [(2m)!!]^2} \nonumber \\
&& \hspace{-0.8cm} \times \left(\frac{6K^2I_2(q)}{kq\mu_{Kk}}\right)^{\ell}
\left(\frac{2kq(\mu_{Kk}^2\!-\!1)}{\mu_{Kk}}\right)^m \, j_{\ell+m}(kq\mu_{Kk}) ,
\label{Psjn}
\eeqa
with
\beq
\mu_k= \frac{\vk\cdot\ve_z}{k} , \;\; 
\mu_{Kk}=\frac{\vK\cdot\vk}{K k} = \frac{1+f\mu_k^2}{\sqrt{1+\mu_k^2(2f+f^2)}} ,
\eeq
\beq
K^2= k^2 [1+\mu_k^2(2f+f^2)] .
\eeq
Expression (\ref{Psjn}) holds for any wavenumber $\vk$ and shows how the
redshift-space power spectrum depends on both $k$ and $\mu_k$.
As recalled above and as is obvious from Eq.(\ref{sd-perp}), the redshift-space power
spectrum for wavevectors perpendicular to the line of sight is equal to the real-space
power spectrum,
\beq
P^s_{\perp}(k_{\perp}) \equiv P^s(\vk_{\perp}) = P(k_{\perp}) ,
\label{Ps-perp}
\eeq
and we can check that we recover Eq.(\ref{Pkjn}) from Eq.(\ref{Psjn}) for $\mu_k=0$.
For longitudinal wavevectors, we simply have $\vK=(1+f)\vk$, so that Eq.(\ref{Psk-Iq})
simplifies as
\beqa
\lefteqn{ \mbox{for } \vk= k_{\parallel} \, \ve_z : \;\;\; P^s_{\parallel}(k_{\parallel})
\equiv P^s(k_{\parallel}\ve_z) } \nonumber \\
&& \hspace{0.3cm} = \int\frac{\dd\vq}{(2\pi)^3} \, e^{\ii\vk\cdot\vq} \,
e^{-k^2(1+f)^2[\sigma_v^2-I_0(q)-(1-3\mu^2)I_2(q)]} ,
\label{Psk-Iq-parallel}
\eeqa
where $\mu=(\vk\cdot\vq)/(k q)=q_{\parallel}/q$, as in Eq.(\ref{PkIn}).
Thus we see at once that, along the line of sight, the effect of redshift distortions
is merely to multiply the amplitude of the linear power spectrum by a factor
$(1+f)^2$; that is, we can absorb the factor $(1+f)^2$ into $\sigma_v^2$, $I_0$ and
$I_2$, or more simply into $P_L(k)$, as was obvious from Eq.(\ref{sd-perp}).
Therefore, we directly obtain from Eq.(\ref{Pkjn}) the expression
\beqa
P^s_{\parallel}(k) & = & \int_0^{\infty}\frac{\dd q \, q^2}{2\pi^2} \,
e^{-k^2(1+f)^2[\sigma_v^2-I_0(q)+2I_2(q)]} \nonumber \\
&& \times \sum_{\ell=0}^{\infty} \left(\frac{6k(1+f)^2I_2(q)}{q}\right)^{\ell}
\, j_{\ell}(kq) ,
\label{Pskjn}
\eeqa
since we only need to multiply each term that involves $P_L(k)$ by a factor
$(1+f)^2$.
In a similar fashion, the standard perturbative expansion (\ref{Pstd}) becomes
\beq
P^s_{\parallel}(k) = \sum_{n=1}^{\infty} P^{s(n)}_{\parallel}(k) \;\; \mbox{with} \;\;
P^{s(n)}_{\parallel}(k) = (1+f)^{2n} \, P^{(n)}(k) ,
\label{sPstd}
\eeq
whereas the ``renormalized'' perturbative expansion (\ref{Psigv}) becomes
\beq
P^s_{\parallel}(k) =  e^{-k^2(1+f)^2\sigma_v^2} \sum_{n=1}^{\infty}
P_{\parallel\sigma_v}^{s(n)}(k) ,
\label{sPsigv}
\eeq
\beq
\mbox{with } \;\;  P_{\parallel\sigma_v}^{s(n)}(k) = (1+f)^{2n} \, P_{\sigma_v}^{(n)}(k) .
\label{sPsigvn}
\eeq
Thus, each order $n$ of the perturbative expansions gets multiplied by a factor
$(1+f)^{2n}$ as we go from real space to redshift space, for modes that are
aligned with the line of sight. In particular, at lowest order we have 
$P^s_{\parallel}(k) = (1+f)^2 P_L(k)+...$, so that we recover the well-known boost factor
of \citet{Kaiser1987}, associated with the infall of galaxies within gravitational potential
wells. On the other hand, the factor $(1+f)^2$ also sharpens the Gaussian damping
prefactor in the expansion (\ref{sPsigv}).

\subsection{``Sticky model'' nonperturbative correction}
\label{Sticky-model-redshift}

\begin{figure*}
\begin{center}
\epsfxsize=9 cm \epsfysize=7 cm {\epsfbox{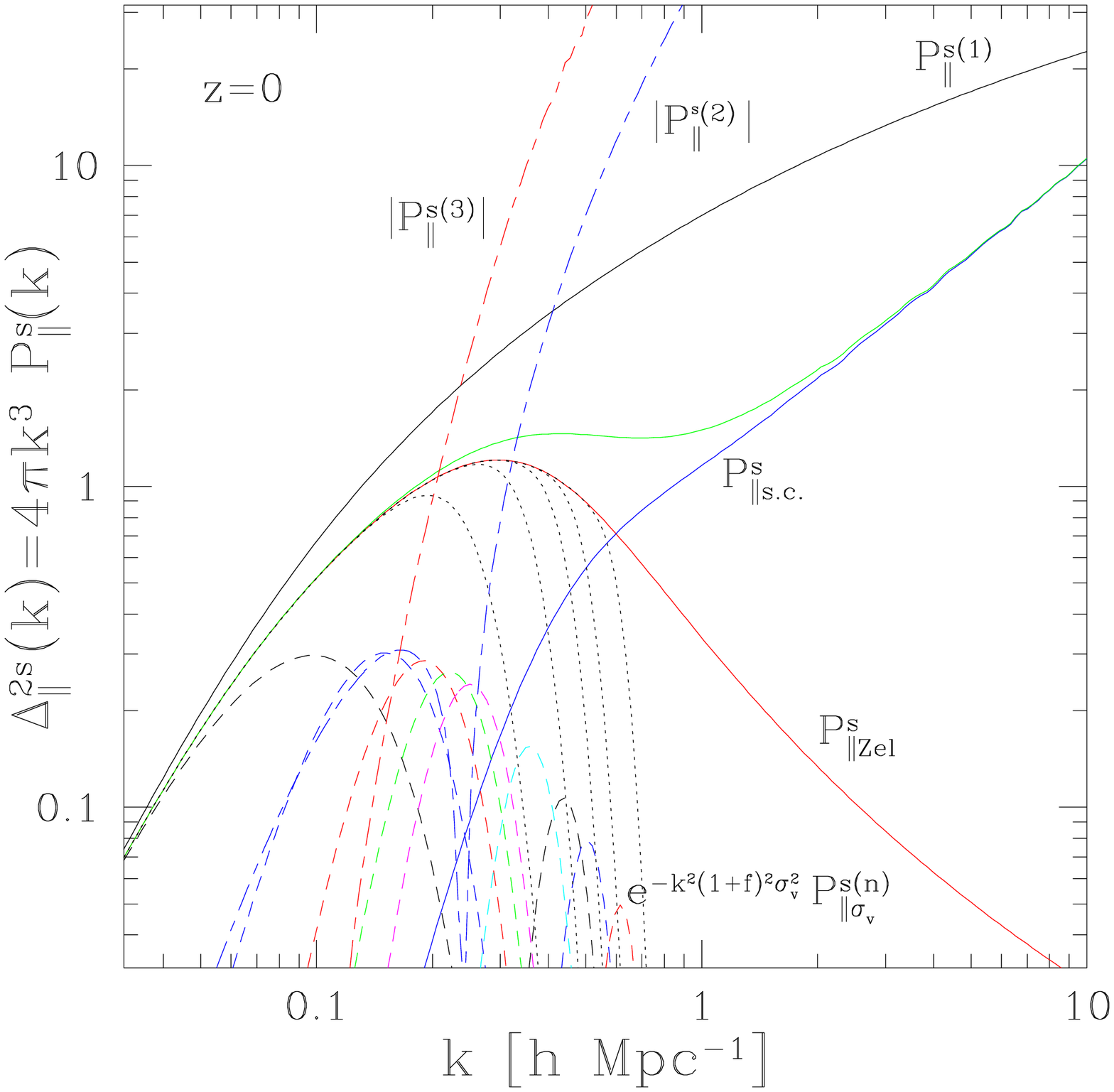}}
\epsfxsize=9 cm \epsfysize=7 cm {\epsfbox{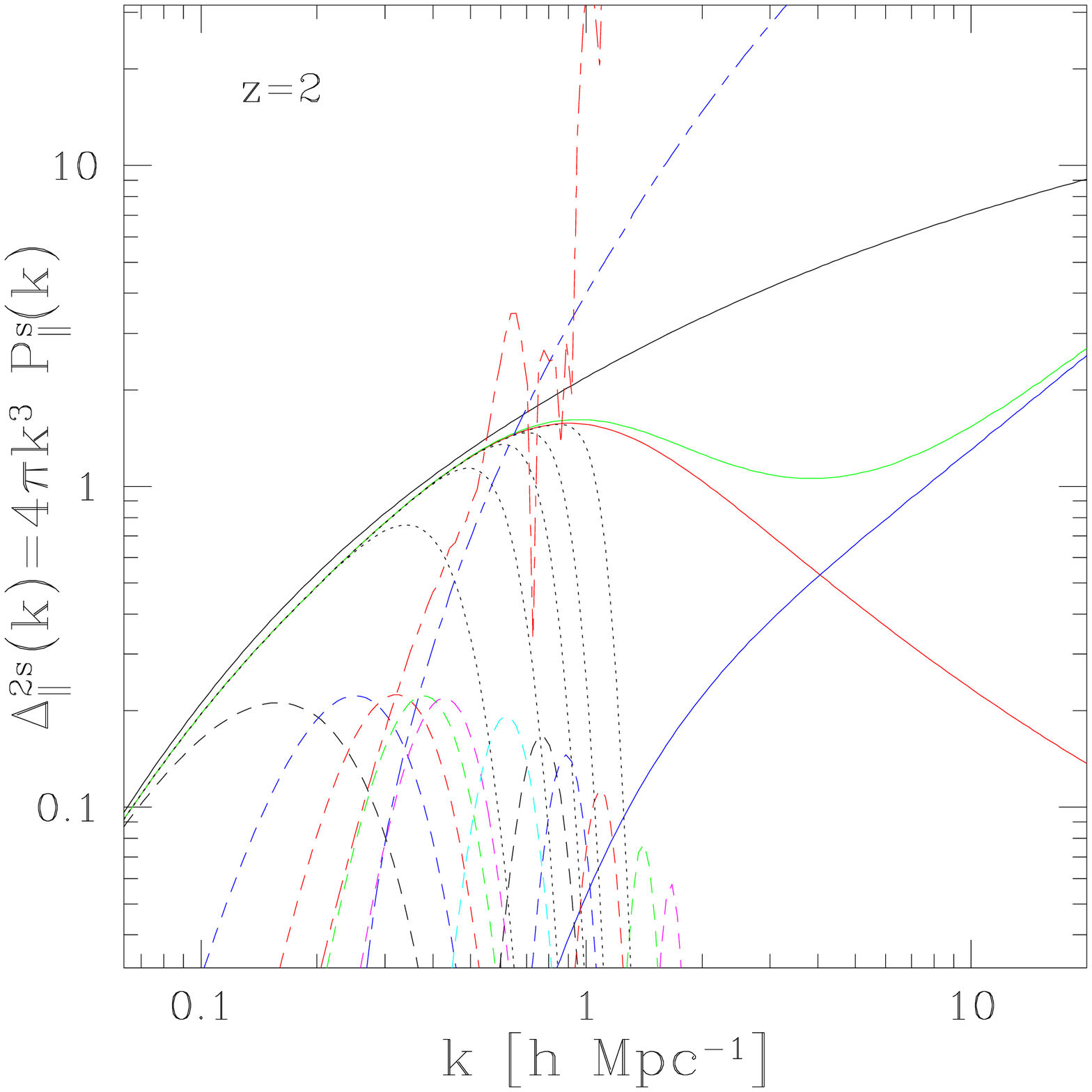}}
\end{center}
\caption{The logarithmic redshift-space power spectrum,
$\Delta^{2s}_{\parallel}(k)= 4\pi k^3 P^s_{\parallel}(k)$, for wavenumbers along the
line of sight. Both plots, at $z=0$ (left panel) and $z=2$ (right panel), are the
redshift-space counterparts of those shown in Fig.~\ref{fig_lDk} for the real-space
power spectrum.}
\label{fig_lsDk}
\end{figure*}

We now turn to the nonperturbative correction associated with the ``sticky model''
introduced in Sect.~\ref{nonperturbative}. Again we focus on the redshift-space
power spectrum $P^s_{\parallel}(k)$ for wavenumbers that are parallel to the 
line of sight. From Eq.(\ref{Pkq-s}) we now need to compute the average
$\lag e^{\ii\vk\cdot\Delta\vs}\rag_{\vq}$.
Before shell crossing, or more precisely, as long as $\Delta x_{L1}>0$ for a
Lagrangian-space separation vector along the $\ve_1$-axis, the ``sticky model''
follows the Zeldovich dynamics,
\beqa
\vk= k \, \ve_z , && \vq= q \, \ve_1 : \nonumber \\
&& \hspace{-1cm} \mbox{if } \; \Delta\Psi_{L1} > -q : \;\;
\vk\cdot\Delta\vs = \vk\cdot\vq + (1+f) \vk\cdot\vPsi_L ,
\label{ks-sticky1}
\eeqa
whereas once $ \Delta x_{L1}<0$ we affect the same position, $\Delta x_1=0$, and the same
peculiar velocity, $\Delta v_1=0$, along the $\ve_1$-axis, for both particles as in
(\ref{Deltax-def}), whence
\beqa
\mbox{if } \; \Delta\Psi_{L1} < -q & : & \Delta s_1 =0 , \mbox{ hence} \nonumber \\
&& \vk\cdot\Delta\vs = (1+f) [ k_2 \Psi_{L2} + k_3 \Psi_{L3} ] .
\label{ks-sticky2}
\eeqa
(The line-of-sight axis $\ve_z$ and the axis $\ve_1$ of the Lagrangian
separation vector $\vq$ are not related.)
At the shell-crossing time we have a discontinuity in the redshift-space
separation $\Delta\vs$, since $\Delta s_1$ jumps from $-f q$ to zero.
Indeed, just before contact, the two particles move closer, with the finite relative
velocity $\Delta v_1=-(a\dot{D}/D)q$, while after collision their relative velocity is
set to zero. This should actually be understood in a loose sense, since within the
``sticky model'' we consider neither transverse directions nor the cloud-in-cloud problem.
Then, as in Sect.~\ref{nonperturbative} the redshift-space power spectrum of the
``sticky model'' reads as
\beq
\Psticky^s(\vk) = \PZel^s(\vk) + \Psc^s(\vk) ,
\label{Ps-sticky-def}
\eeq
and from Eqs.(\ref{ks-sticky1})-(\ref{ks-sticky2}) we obtain for wavenumbers along the
line of sight
\beqa
\Pscpar(k) \!\! & = & \!\!\! \int \! \frac{\dd q \, q^2}{(2\pi)^2} \,
e^{-q^2/(2\sigma_{\parallel}^2)} \! \int_0^1 \!\!\! \dd\mu \,
e^{-k^2(1+f)^2(1-\mu^2) [\sigma_v^2-I_0-I_2]}
\nonumber \\
&& \hspace{-1.3cm} \times \Real \left\{
w\!\left(\!\frac{\ii q}{\sqrt{2}\sigma_{\parallel}}\!\right) -
e^{-\ii k f q \mu} \,
w\!\left(\!\frac{\ii q-k(1\!+\!f)\mu\sigma_{\parallel}^2}
{\sqrt{2}\sigma_{\parallel}}\right)\!\right\} .
\label{Psc-par}
\eeqa
As compared with Eq.(\ref{Psc-def1}), the new exponential factor $e^{-\ii k f q \mu}$
is due to the discontinuity of $\Delta\vs$ at $\Delta\Psi_{L1} = -q$, associated with
shell crossing within the ``sticky model''. The factors $(1+f)$ that multiply the
longitudinal wavenumber $k$ could be expected from Eqs.(\ref{ks-sticky1})-(\ref{ks-sticky2}).

\subsection{Redshift-space logarithmic power}
\label{Redshift-space-logarithmic-power}

\begin{figure*}
\begin{center}
\epsfxsize=8.5 cm \epsfysize=6 cm {\epsfbox{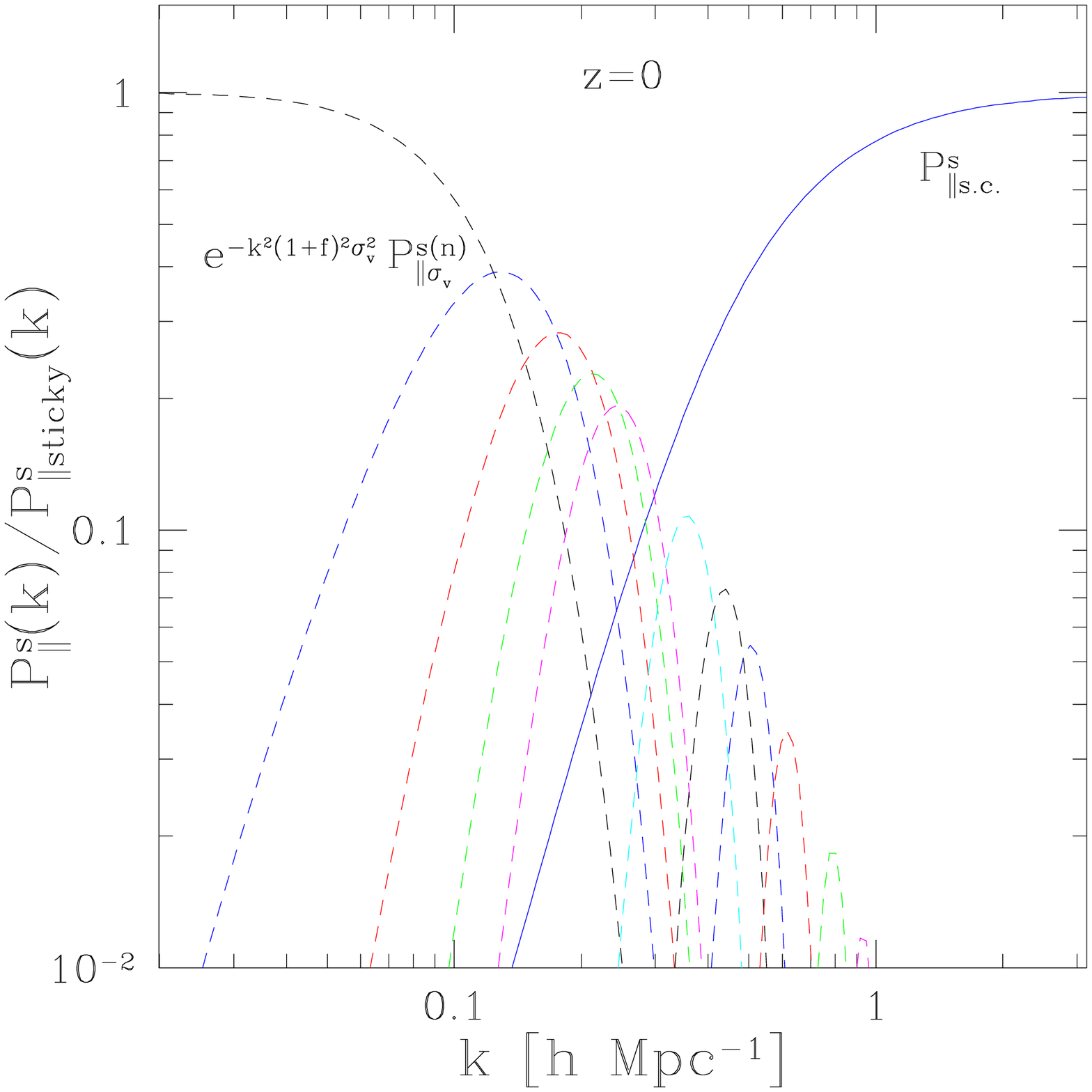}}
\epsfxsize=8.5 cm \epsfysize=6 cm {\epsfbox{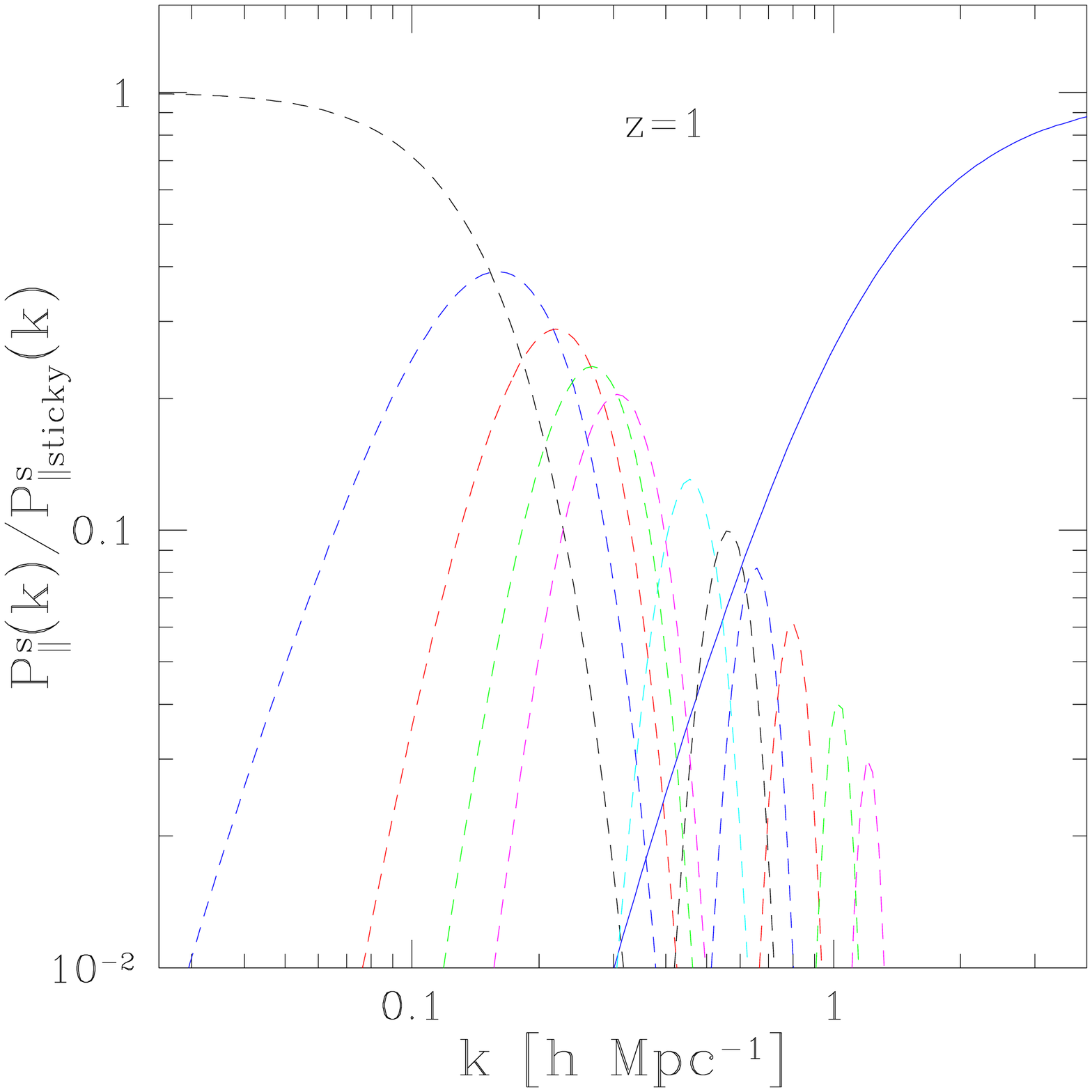}}\\
\epsfxsize=8.5 cm \epsfysize=6 cm {\epsfbox{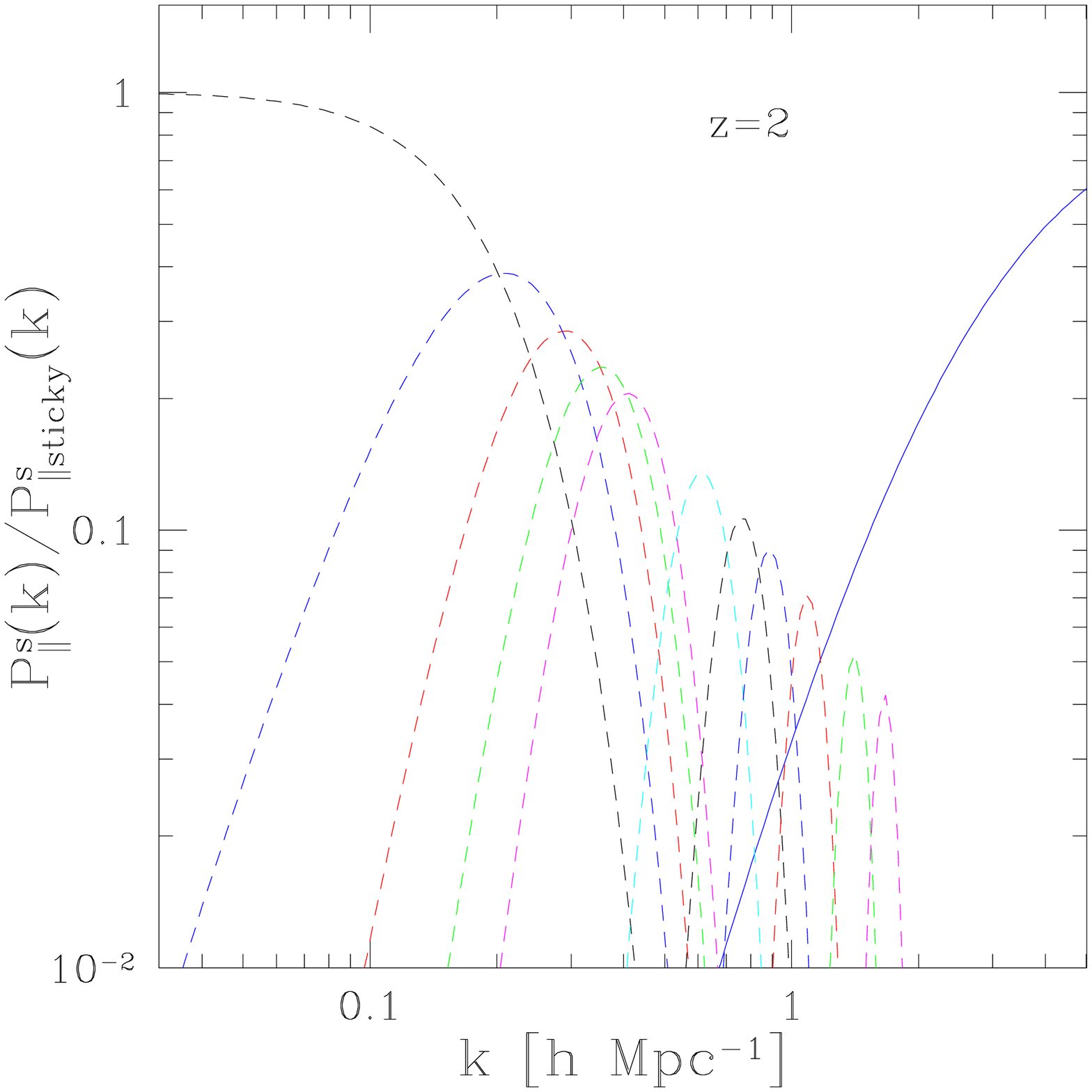}}
\epsfxsize=8.5 cm \epsfysize=6 cm {\epsfbox{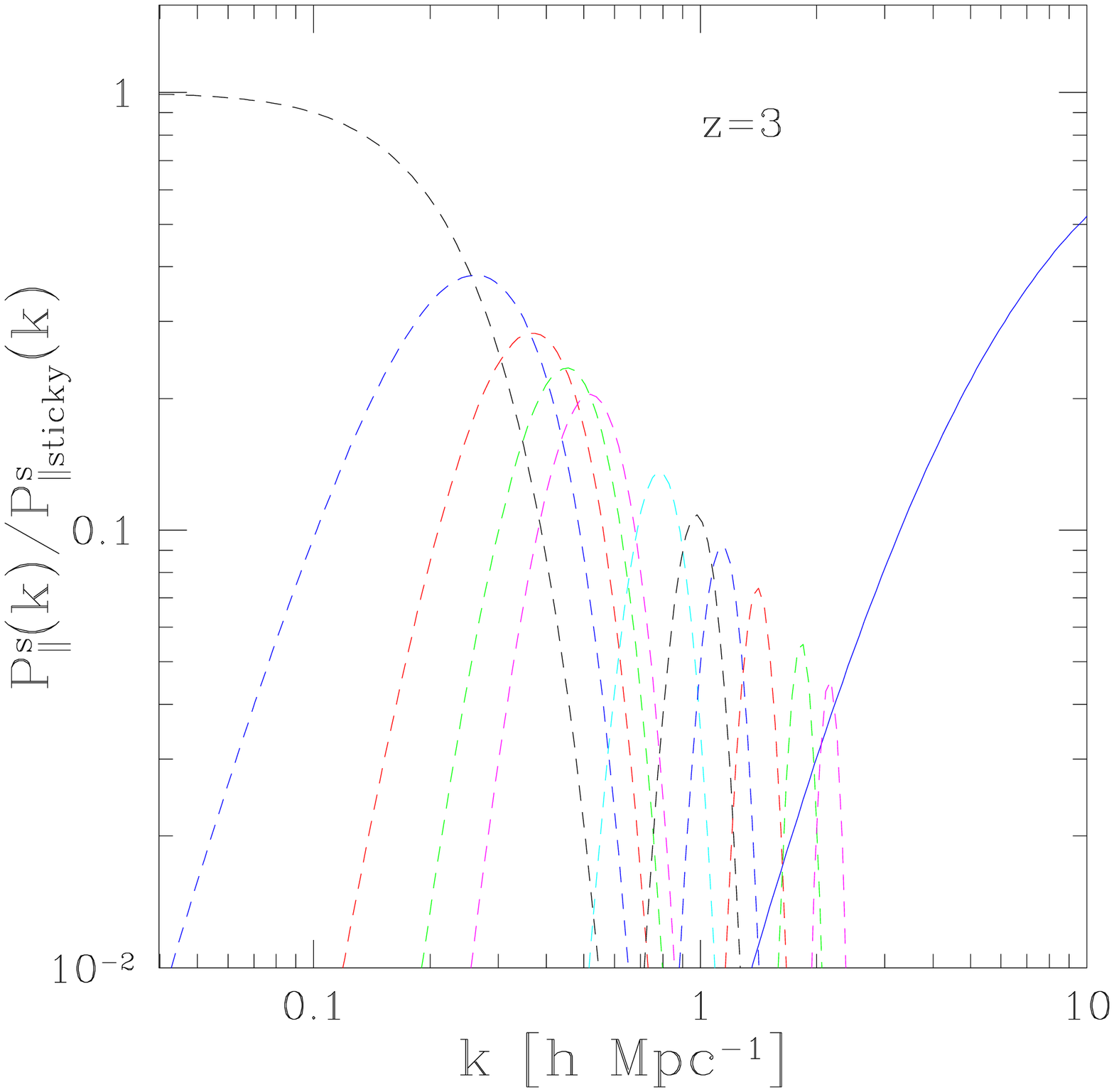}}
\end{center}
\caption{The ratios of the ``renormalized'' perturbative terms
of the expansion (\ref{sPsigv}) to the nonlinear ``sticky model'' redshift-space power
spectrum, and of the  nonperturbative contribution $\Pscpar(k)$. This is the
redshift-space counterpart of the real-space Fig.~\ref{fig_lrDk}, for longitudinal
wavenumbers.}
\label{fig_lrsDk}
\end{figure*}

As for the real-space Fig.~\ref{fig_lDk}, we show in Fig.~\ref{fig_lsDk} our numerical
results for the redshift-space logarithmic power, for longitudinal wavenumbers
$\vk$, defined as
\beq
\Delta^{2s}_{\parallel}(k) = 4\pi k^3 P^s_{\parallel}(k) .
\label{sDelta2def}
\eeq
We use the same definition (\ref{Delta2def}), even though $P^s_{\parallel}(k)$ only
holds along the longitudinal direction and $P^s_{\perp}(k) (=P(k))$ holds along the
two transverse directions (so that using a factor $k$ instead of $k^3$ would be
more natural here), to make the comparison with Fig.~\ref{fig_lDk} easier.
In particular, this means that $\Delta^{2s}_{\perp}(k)$ is given by Fig.~\ref{fig_lDk}
for transverse wavenumbers.

As compared with Fig.~\ref{fig_lDk}, we can check that the linear-regime power,
$P^{s(1)}_{\parallel}$, is amplified by a factor $(1+f)^2$, whereas standard higher order
terms are amplified by factors $(1+f)^{2n}$, see Eq.(\ref{sPstd}). 
Even though the same factors apply to the higher order terms of the ``renormalized''
perturbative expansion, see Eq.(\ref{sPsigvn}), their peak height is not greater than in
Fig.~\ref{fig_lDk} for $n\geq 2$ because of the stronger Gaussian damping prefactor
$e^{-k^2(1+f)^2\sigma_v^2}$ in Eq.(\ref{sPsigv}). For large $n$ this even leads to a smaller
amplitude as compared with the associated real-space contribution.
This makes the full nonlinear Zeldovich and ``sticky model'' power spectra greater
than their
real-space counterparts on large scales, in the quasi-linear regime, but smaller in the
highly nonlinear regime. In particular, it is clear from Eq.(\ref{Psk-Iq-parallel}) that
the high-$k$ damping of the Zeldovich power spectrum becomes sharper because of the
factor $(1+f)^2$ in the exponential term (even though this only leads to a power-law
decay as noticed in Sect.~\ref{Logarithmic-power}).
As well as for the real-space power spectrum, the ``renormalized'' perturbative
expansion (\ref{sPsigv}) is more convenient to distinguish the relative contributions
of higher order terms, as they are all positive and do not show the cancellations
associated with the standard expansion (\ref{sPstd}).

Since within the ``sticky model'' we have after shell crossing $\Delta s_1=0$, just as we
had $\Delta x_1=0$ in real space, we again obtain a $k^{-2}$ tail at high $k$
for the nonperturbative term $\Pscpar(k)$, whence
$\Delta^{2s}_{\parallel \rm sticky}(k) \sim k$, as would be the case for planar
structures in redshift space.
However, because the dominant effect of redshift distortions on small scales is to
decrease the power, through damping factors of the form $e^{-k^2(1+f)^2\sigma_v^2}$,
the nonperturbative redshift-space contribution $\Pscpar(k)$ is less than its
real-space counterpart. Coupled with the faster decay of $\PZelpar(k)$, this leads to
a temporary decrease of $\Delta^{2s}_{\parallel \rm sticky}(k)$ at $k\sim 1$ to
$4 h$ Mpc$^{-1}$ at $z=2$, before the asymptotic tail $\propto k$ becomes
dominant.

The ``sticky model'' is not intended here to describe
the redshift-space power spectrum better than the Zeldovich approximation.
Indeed, on small scales setting $\Delta v_1=0$ is not realistic, since one should
rather describe multi-streaming and take into account the finite velocity dispersion
of collapsed objects built by virialization processes. Then, instead of a ``sticking''
in redshift space one obtains a finite size extension that is greater than the
real-space size of the object (because of the additional term associated with the
velocity dispersion in Eq.(\ref{sdef})). As is known, this explains the
characteristic ``fingers of god'' observed in galaxy surveys 
\citep{Jackson1972,Tadros1999}, and instead of
a universal growing tail $\Delta^{2s}_{\parallel \rm sticky}(k)\propto k$, this leads to
a decaying tail (as for the Zeldovich power spectrum but with different quantitative
properties, since the latter are set by the complex nonlinear virialization processes
rather than the linear velocity field).
Of course, it is possible to modify the ``sticky model'' to implement such a
velocity dispersion and to obtain more realistic results in redshift space.
For instance, while retaining ``sticking'' in real space, one could allocate a random
relative velocity to the pair, so that the perturbative expansions remain
unchanged and the nonperturbative contribution involves a damping term
of the form $e^{-k^2\sigma_{\rm vir}^2}$, where $\sigma_{\rm vir}$ is the
characteristic velocity dispersion in virialized objects.
However, this goes beyond the scope of this article, since here we are only
interested in the comparison of perturbative and nonperturbative terms
and the simple ``sticky model'' in the form (\ref{ks-sticky2}) is sufficient to
estimate the scale where nonperturbative corrections become important.

\subsection{Redshift-space perturbative and nonperturbative contributions}
\label{Redshift-space-corrections}

Finally, as for the real-space Fig.~\ref{fig_lrDk}, we show in Fig.~\ref{fig_lrsDk}
the ratios of the successive terms of the ``renormalized'' perturbative expansion
(\ref{sPsigv}) with respect to the nonlinear power spectrum of the ``sticky model'',
for the same orders $n=1$ to $5$, as well as $n=10, 15, 20, 30, 50$, and $70$.
We also plot the ratio $\Pscpar(k)/P_{\parallel \rm sticky}(k)$, to compare with the
amplitude of the nonperturbative correction associated with shell crossing effects,
and we consider the four redshifts $z=0,1,2$, and $3$.
As for the real-space power spectrum, the potential of perturbative expansions is
greater at higher redshift, as we can include more perturbative terms
before they become subdominant as compared with the nonperturbative contribution
$\Pscpar(k)$, and they extend by a larger factor the range of $k$ where systematic
results can be obtained.

\begin{table}
\begin{center}
\begin{tabular}{c||c|c|c||c}
$z$ & $\%$ & $k^s_{\parallel 1\rm loop} \; [h {\rm Mpc}^{-1}]$ & $k^s_{\parallel \rm s.c.} \; [h {\rm Mpc}^{-1}]$ & $n^s_{\parallel \rm s.c.}$  \rule[-0.3cm]{0cm}{0.6cm} \\ \hline\hline
 & $1\%$ & 0.025 & 0.13 &  \rule[-0.2cm]{0cm}{0.6cm} \\
0 & $10\%$ & 0.052 & 0.26 & 8 \rule[-0.2cm]{0cm}{0.4cm} \\ 
 & $50\%$ & & 0.57 &  \rule[-0.2cm]{0cm}{0.4cm} \\ \hline
 & $1\%$ & 0.03 & 0.3 &  \rule[-0.2cm]{0cm}{0.6cm} \\
1 & $10\%$ & 0.064 & 0.63 & 18 \rule[-0.2cm]{0cm}{0.4cm} \\
 & $50\%$ & & 1.3 &  \rule[-0.2cm]{0cm}{0.4cm} \\ \hline
 & $1\%$ & 0.033 & 0.68 & \rule[-0.2cm]{0cm}{0.6cm}  \\
2 & $10\%$ & 0.08 & 1.5 & 37 \rule[-0.2cm]{0cm}{0.4cm} \\
 & $50\%$ & & 4 &  \rule[-0.2cm]{0cm}{0.4cm} \\ \hline
 & $1\%$ & 0.043 & 1.3 & \rule[-0.2cm]{0cm}{0.6cm} \\
3 & $10\%$ & 0.1 & 3.2 & 72 \rule[-0.2cm]{0cm}{0.4cm} \\ 
 & $50\%$ & & 9.8 &  \rule[-0.2cm]{0cm}{0.4cm} \\
\end{tabular}
\end{center}
\caption{The dependence on redshift $z$ of the wavenumbers where the ratios
$e^{-k^2(1+f)^2\sigma_v^2} P_{\parallel\sigma_v}^{s(2)}(k)/P^s_{\parallel \rm sticky}(k)$
(at $k^s_{\parallel 1\rm loop}$) and
$\Pscpar(k)/P^s_{\parallel \rm sticky}(k)$ (at $k^s_{\parallel \rm s.c.}$) reach $1\%$,
$10\%$, and $50\%$.
The last column gives the last order $n^s_{\parallel \rm s.c.}$ of the ``renormalized''
perturbative expansion that is not fully below the nonperturbative term.
This table is the redshift-space counterpart of the real-space Table~\ref{Table_range},
for longitudinal wavenumbers.}
\label{sTable_range}
\end{table}

We give in Table~\ref{sTable_range} the wavenumbers $k^s_{\parallel 1\rm loop}$
and $k^s_{\parallel \rm s.c.}$ where the second order of perturbation theory,
$P_{\parallel\sigma_v}^{s(2)}(k)$, and the nonperturbative contribution, $\Pscpar(k)$,
reach levels of $1\%$, $10\%$, or $50\%$.
As noticed above from the comparison of Figs.~\ref{fig_lrsDk} and \ref{fig_lrDk},
the interval $[k^s_{\parallel 1\rm loop},k^s_{\parallel \rm s.c.}]$, where higher orders
of perturbation theory apply, shifts to higher $k$ and becomes broader at higher
redshift, as for the real-space spectrum described in Table~\ref{Table_range}. However,
it is shifted towards smaller $k$ than with the real space power spectrum,
as could be expected from the fact that redshift-space
perturbative terms of order $n$ are multiplied by prefactors $(1+f)^{2n}$, see
Eqs.(\ref{sPstd})-(\ref{sPsigvn}), so that higher order terms become relevant earlier
and on larger scales.
On the other hand, the order $n^s_{\parallel \rm s.c.}$, after which perturbative
contributions become subdominant as compared with the nonperturbative contribution,
is similar to the one obtained for the real-space power spectrum, and grows from
$8$ at $z=0$ up to $72$ at $z=3$.

Since redshift distortions are greatest along the line of sight and they vanish
for transverse wavenumbers, one would obtain similar results for multipoles
of the redshift-space power spectrum, where one expands the dependence
on the angle with respect to the line of sight of $P^s(\vk)$ over
Legendre polynomials of $\mu=(\vk\cdot\ve_z)/k$. On a quantitative level,
one would obtain results in-between those presented in Tables~\ref{Table_range} 
and \ref{sTable_range}.

\section{Conclusion}
\label{Conclusion}

In this article we have introduced a ``sticky model'' that coincides with the
usual Zeldovich dynamics before shell crossing, while after shell crossing it includes
a sticking of particle pairs along their longitudinal direction.
This implies that the nonlinear density power spectra of both models have the
same perturbative expansions and only differ by nonperturbative terms
that arise from the dynamics beyond shell crossing.
Since we can obtain explicit expressions for perturbative terms at all orders and for
this nonperturbative correction, we have been able to compare their respective
amplitudes in detail, focusing on a $\Lambda$CDM cosmology.

In real space, we find that perturbative approaches based on the fluid description
can extend the wavenumber up to which systematic analytical predictions can be
obtained from $k\simeq 0.033 h$ Mpc$^{-1}$ up to $\simeq 0.23 h$ Mpc$^{-1}$
at $z=0$, and from $\simeq 0.07 h$ Mpc$^{-1}$ up to $\simeq 2.2 h$ Mpc$^{-1}$
at $z=3$, as compared with linear theory (if we need an accuracy of $1\%$).
We also give detailed results for other redshifts and for $10\%$ and $50\%$ 
accuracy levels.
Going to higher $k$ requires taking shell-crossing effects into account.
Nevertheless, these results show that such perturbative approaches, based
on hydrodynamical equations of motion, have a significant potential, especially
at high and moderate redshifts. Since it is unlikely that it will be possible
to explicitly compute high-order perturbative terms up to the last order $n_{\rm s.c.}$
which is above the nonperturbative correction ($n_{\rm s.c.}=9$ at $z=0$ and
$n_{\rm s.c.}=66$ at $z=3$), this provides a strong incentive to develop
resummation schemes. Then, the hope is that such methods can efficiently resum
most of the contributions associated with higher order terms and
achieve a faster convergence.

This analysis is also useful in the context of cosmological reconstruction, where
one tries to recover the dynamical history of a given region of the sky from its
present density field (and next to derive from this reconstruction the present velocity
field). Indeed, these methods usually neglect shell-crossing effects.
In particular, we find that the Monge-Amp\`ere-Kantorovich method is close to optimal
at $z=0$, as it fares well down to the scale where shell-crossing contributions
dominate.
With respect to the reconstruction of the baryon acoustic peak or of the acoustic
oscillations of the power spectrum, there seems to be only modest room for
improvement over current methods based on the linear displacement field at $z=0$,
in agreement with only a few orders of perturbation theory being relevant
($n_{\rm s.c.}=9$). At higher $z$ one could in principle do much better by including
higher order terms, but since baryon acoustic oscillations of the linear power spectrum
have a small amplitude at high $k$, this may not significantly improve the
signal-to-noise ratio. However, this remains useful for other observational probes that
are sensitive to the shape and amplitude of the power spectrum on weakly nonlinear
scales, such as weak-lensing surveys.

We have also pointed out that the behavior of the system after shell crossing 
plays a key role in the shape of the density power spectrum on mildly nonlinear
scales, $1< \Delta^2(k) < 100$. Thus, arguments based on the spherical collapse 
dynamics are not sufficient to explain the steep rise of the nonlinear power
in this intermediate regime. Indeed, while for the Zeldovich dynamics such
a growth is almost entirely wiped out by the escape of particles to infinity
(as the random linear displacements erase small-scale features and lead to a decay
of $\Delta^2(k)$ at high $k$), for the ``sticky model'' there is a regular growth
up to the high-$k$ asymptote, $\Psticky(k) \sim k^{-2}$. 
Therefore, it is not sufficient to consider the spherical collapse of a ``typical''
overdensity to explain the shape of $P(k)$ on these scales, as one must take the
Gaussian average over a whole range of initial density fluctuations, which show
a significant amount of shell crossing that has a strong impact on the resulting
power spectrum.

Finally, we have obtained similar results for the redshift-space power spectrum.
Again, the scope of perturbative approaches (based on the fluid description)
is greater at higher $z$. As compared with the real-space power spectrum,
the characteristic wavenumbers where higher order terms of the perturbative
expansions and nonperturbative contributions come into play are shifted to
lower values of $k$. Indeed, because of the amplification
of perturbations from the uniform Hubble background by the additional contribution
from peculiar velocities, the nonlinear regime is reached on larger scales than in real
space.

The ``sticky model'' presented in this article, and more
generally the expression (\ref{Pkxq}) of the density power spectrum in terms of
the Lagrangian displacement, could serve as a basis for models for the nonlinear
power spectrum. In particular, the spirit of this model could be incorporated
into the Lagrangian-based models, associated with the parametric system
(\ref{kL})-(\ref{f-def}), or into the ``halo model'', to make the bridge between the
perturbative regime and the highly nonlinear, nonperturbative regime
\citep{Valageas2010a}.
This could also be extended to redshift space by including a model for the
velocity dispersion within virialized objects.
However, this is beyond the scope of the present article so we leave it
to future works.

\bibliographystyle{aa} % style aa.bst
\bibliography{15658}

\begin{thebibliography}{65}
\expandafter\ifx\csname natexlab\endcsname\relax\def\natexlab#1{#1}\fi

\bibitem[{Abramowitz \& Stegun(1970)}]{Abramowitz}
Abramowitz, M. \& Stegun, I.~A. 1970, Handbook of Mathematical Functions (New
  York: Dover)

\bibitem[{Afshordi(2007)}]{Afshordi2007}
Afshordi, N. 2007, Phys. Rev. D, 75, 021302

\bibitem[{Alcock \& Paczynski(1979)}]{Alcock1979}
Alcock, C. \& Paczynski, B. 1979, Nature, 281, 358

\bibitem[{Bec \& Khanin(2007)}]{Bec2007}
Bec, J. \& Khanin, K. 2007, Phys. Rep., 447, 1

\bibitem[{Bernardeau {et~al.}(2002)Bernardeau, Colombi, {Gazta\~naga}, \&
  Scoccimarro}]{Bernardeau2002}
Bernardeau, F., Colombi, S., {Gazta\~naga}, E., \& Scoccimarro, R. 2002, Phys.
  Rep., 367, 1

\bibitem[{Bernardeau \& Kofman(1995)}]{Bernardeau1995}
Bernardeau, F. \& Kofman, L. 1995, Astrophys. J., 443, 479

\bibitem[{Bernardeau \& Valageas(2008)}]{BernardeauVal2008}
Bernardeau, F. \& Valageas, P. 2008, Phys. Rev. D, 78, 083503

\bibitem[{Bernardeau \& Valageas(2010{\natexlab{a}})}]{BernardeauVal2010a}
Bernardeau, F. \& Valageas, P. 2010{\natexlab{a}}, Phys. Rev. D, 81, 043516

\bibitem[{Bernardeau \& Valageas(2010{\natexlab{b}})}]{BernardeauVal2010b}
Bernardeau, F. \& Valageas, P. 2010{\natexlab{b}}, Phys. Rev. E

\bibitem[{Bond {et~al.}(1991)Bond, Cole, Efstathiou, \& Kaiser}]{Bond1991}
Bond, J.~R., Cole, S., Efstathiou, G., \& Kaiser, N. 1991, Astrophys. J., 379,
  440

\bibitem[{Brenier {et~al.}(2003)Brenier, Frisch, Henon, Loeper, Matarrese,
  Mohayaee, \& Sobolevskii}]{Brenier2003}
Brenier, Y., Frisch, U., Henon, M., {et~al.} 2003, Mon. Not. R. Astron. Soc.,
  346, 501

\bibitem[{Buchert(1994)}]{Buchert1994}
Buchert, T. 1994, Mon. Not. R. Astron. Soc., 267, 811

\bibitem[{Carlson {et~al.}(2009)Carlson, White, \& Padmanabhan}]{Carlson2009}
Carlson, J., White, M., \& Padmanabhan, N. 2009, Phys. Rev. D, 80, 043531

\bibitem[{Coles {et~al.}(1993)Coles, Melott, \& Shandarin}]{Coles1993}
Coles, P., Melott, A.~L., \& Shandarin, S.~F. 1993, Mon. Not. R. Astron. Soc.,
  260, 765

\bibitem[{Cooray \& Sheth(2002)}]{Cooray2002}
Cooray, A. \& Sheth, R. 2002, Phys. Rep., 372, 1

\bibitem[{Crocce \& Scoccimarro(2006{\natexlab{a}})}]{Crocce2006b}
Crocce, M. \& Scoccimarro, R. 2006{\natexlab{a}}, Phys. Rev. D, 73, 063520

\bibitem[{Crocce \& Scoccimarro(2006{\natexlab{b}})}]{Crocce2006a}
Crocce, M. \& Scoccimarro, R. 2006{\natexlab{b}}, Phys. Rev. D, 73, 063519

\bibitem[{Crocce \& Scoccimarro(2008)}]{Crocce2008}
Crocce, M. \& Scoccimarro, R. 2008, Phys. Rev. D, 77, 023533

\bibitem[{Eisenstein {et~al.}(1998)Eisenstein, Hu, \& Tegmark}]{Eisenstein1998}
Eisenstein, D.~J., Hu, W., \& Tegmark, M. 1998, Astrophys. J. Lett., 504, 57

\bibitem[{Eisenstein {et~al.}(2007)Eisenstein, Seo, Sirko, \&
  Spergel}]{Eisenstein2007}
Eisenstein, D.~J., Seo, H.-J., Sirko, E., \& Spergel, D.~N. 2007, Astrophys.
  J., 664, 675

\bibitem[{Eisenstein {et~al.}(2005)Eisenstein, Zehavi, \&
  et~al.}]{Eisenstein2005}
Eisenstein, D.~J., Zehavi, I., \& et~al., D. W.~H. 2005, Astrophys. J., 633,
  560

\bibitem[{Goroff {et~al.}(1986)Goroff, Grinstein, Rey, \& Wise}]{Goroff1986}
Goroff, M.~H., Grinstein, B., Rey, S.-J., \& Wise, M.~B. 1986, Astrophys. J.,
  311, 6

\bibitem[{Gurbatov {et~al.}(1991)Gurbatov, Malakhov, \& Saichev}]{Gurbatov1991}
Gurbatov, S., Malakhov, A., \& Saichev, A. 1991, Nonlinear random waves and
  turbulence in nondispersive media: waves, rays and particles (Manchester
  University Press)

\bibitem[{Gurbatov {et~al.}(1989)Gurbatov, Saichev, \&
  Shandarin}]{Gurbatov1989}
Gurbatov, S.~N., Saichev, A.~I., \& Shandarin, S.~F. 1989, Mon. Not. R. Astron.
  Soc., 236, 385

\bibitem[{Hamilton(1992)}]{Hamilton1992}
Hamilton, A. J.~S. 1992, Astrophys. J. Lett., 385, 5

\bibitem[{Hamilton {et~al.}(1991)Hamilton, Kumar, Lu, \&
  Matthews}]{Hamilton1991}
Hamilton, A. J.~S., Kumar, P., Lu, E., \& Matthews, A. 1991, Astrophys. J.
  Lett., 374, 1

\bibitem[{Jackson(1972)}]{Jackson1972}
Jackson, J.~C. 1972, Mon. Not. R. Astron. Soc., 156, 1P

\bibitem[{Kaiser(1987)}]{Kaiser1987}
Kaiser, N. 1987, Mon. Not. R. Astron. Soc., 227, 1

\bibitem[{Kraichnan(1959)}]{Kraichnan1959}
Kraichnan, R.~H. 1959, J. Fluid Mech., 5, 497

\bibitem[{Massey {et~al.}(2007)Massey, Rhodes, \& et~al.}]{Massey2007}
Massey, R., Rhodes, J., \& et~al., A.~L. 2007, Astrophys. J. Supp., 172, 239

\bibitem[{Matarrese \& Pietroni(2007)}]{Matarrese2007}
Matarrese, S. \& Pietroni, M. 2007, JCAP, 6, 26

\bibitem[{Matsubara(2008)}]{Matsubara2008}
Matsubara, T. 2008, Phys. Rev. D, 77, 063530

\bibitem[{McDonald(2007)}]{McDonald2007}
McDonald, P. 2007, Phys. Rev. D, 75, 043514

\bibitem[{Melott {et~al.}(1994)Melott, Shandarin, \& Weinberg}]{Melott1994}
Melott, A.~L., Shandarin, S.~F., \& Weinberg, D.~H. 1994, Astrophys. J., 428,
  28

\bibitem[{Mohayaee {et~al.}(2006)Mohayaee, Mathis, Colombi, \&
  Silk}]{Mohayaee2006}
Mohayaee, R., Mathis, H., Colombi, S., \& Silk, J. 2006, Mon. Not. R. Astron.
  Soc., 365, 939

\bibitem[{Munshi {et~al.}(2008)Munshi, Valageas, van Waerbeke, \&
  Heavens}]{Munshi2008}
Munshi, D., Valageas, P., van Waerbeke, L., \& Heavens, A. 2008, Phys. Rep.,
  462, 67

\bibitem[{Navarro {et~al.}(1997)Navarro, Frenk, \& White}]{NFW1997}
Navarro, J.~F., Frenk, C.~S., \& White, S. D.~M. 1997, Astrophys. J., 490, 493

\bibitem[{Noullez \& Vergassola(1994)}]{Noullez1994}
Noullez, A. \& Vergassola, M. 1994, J. Sci. Comput., 9, 259

\bibitem[{Padmanabhan(1996)}]{Padmanabhan1996}
Padmanabhan, T. 1996, Mon. Not. R. Astron. Soc., 278, L29

\bibitem[{Pauls \& Melott(1995)}]{Pauls1995}
Pauls, J.~L. \& Melott, A.~L. 1995, Mon. Not. R. Astron. Soc., 274, 99

\bibitem[{Peacock \& Dodds(1996)}]{Peacock1996}
Peacock, J.~A. \& Dodds, S.~J. 1996, Mon. Not. R. Astron. Soc., 280, L19

\bibitem[{Peebles(1974)}]{Peebles1974}
Peebles, P. J.~E. 1974, Astron. Astrophys., 32, 391

\bibitem[{Peebles(1980)}]{Peebles1980}
Peebles, P. J.~E. 1980, The large scale structure of the universe (Princeton:
  Princeton university press)

\bibitem[{Peebles(1982)}]{Peebles1982}
Peebles, P. J.~E. 1982, Astrophys. J. Lett., 263, 1

\bibitem[{Peebles(1989)}]{Peebles1989}
Peebles, P. J.~E. 1989, Astrophys. J. Lett., 344, 53

\bibitem[{Pietroni(2008)}]{Pietroni2008}
Pietroni, M. 2008, JCAP, 10, 36

\bibitem[{Sathyaprakash {et~al.}(1995)Sathyaprakash, Sahni, Munshi, Pogosyan,
  \& Melott}]{Sathyaprakash1995}
Sathyaprakash, B.~S., Sahni, V., Munshi, D., Pogosyan, D., \& Melott, A.~L.
  1995, Mon. Not. R. Astron. Soc., 275, 463

\bibitem[{Schneider \& Bartelmann(1995)}]{Schneider1995}
Schneider, P. \& Bartelmann, M. 1995, Mon. Not. R. Astron. Soc., 273, 475

\bibitem[{Seo {et~al.}(2010)Seo, Eckel, \& et~al.}]{Seo2010}
Seo, H.-J., Eckel, J., \& et~al., D. J.~E. 2010, Astrophys. J.,

\bibitem[{Smith {et~al.}(2003)Smith, Peacock, \& et~al.}]{Smith2003}
Smith, R.~E., Peacock, J.~A., \& et~al., A.~J. 2003, Mon. Not. R. Astron. Soc.,
  341, 1311

\bibitem[{Tadros {et~al.}(1999)Tadros, Ballinger, \& et~al.}]{Tadros1999}
Tadros, H., Ballinger, W.~E., \& et~al., A. N.~T. 1999, Mon. Not. R. Astron.
  Soc., 305, 527

\bibitem[{Taruya \& Hiramatsu(2008)}]{Taruya2008}
Taruya, A. \& Hiramatsu, T. 2008, Astrophys. J., 674, 617

\bibitem[{Taruya {et~al.}(2009)Taruya, Nishimichi, Saito, \&
  Hiramatsu}]{Taruya2009}
Taruya, A., Nishimichi, T., Saito, S., \& Hiramatsu, T. 2009, Phys. Rev. D, 80,
  123503

\bibitem[{Taylor \& Hamilton(1996)}]{Taylor1996}
Taylor, A.~N. \& Hamilton, A. J.~S. 1996, Mon. Not. R. Astron. Soc., 282, 767

\bibitem[{Valageas(2004)}]{Valageas2004}
Valageas, P. 2004, Astron. Astroph., 421, 23

\bibitem[{Valageas(2007{\natexlab{a}})}]{Valageas2007a}
Valageas, P. 2007{\natexlab{a}}, Astron. Astroph., 465, 725

\bibitem[{Valageas(2007{\natexlab{b}})}]{Valageas2007b}
Valageas, P. 2007{\natexlab{b}}, Astron. Astroph., 476, 31

\bibitem[{Valageas(2008)}]{Valageas2008}
Valageas, P. 2008, Astron. Astroph., 484, 79

\bibitem[{Valageas(2009)}]{Valageas2009b}
Valageas, P. 2009, Phys. Rev. E, 80, 016305

\bibitem[{Valageas \& Bernardeau(2010)}]{Valageas2010b}
Valageas, P. \& Bernardeau, F. 2010, arXiv:1009.1974, submitted to Phys. Rev. D

\bibitem[{Valageas \& Nishimichi(2010)}]{Valageas2010a}
Valageas, P. \& Nishimichi, T. 2010, arXiv:1009.0597

\bibitem[{Valageas \& Schaeffer(1997)}]{Valageas1997}
Valageas, P. \& Schaeffer, R. 1997, Astron. Astroph., 328, 435

\bibitem[{Vergassola {et~al.}(1994)Vergassola, Dubrulle, Frisch, \&
  Noullez}]{Vergassola1994}
Vergassola, M., Dubrulle, B., Frisch, U., \& Noullez, A. 1994, Astron.
  Astrophys., 289, 325

\bibitem[{Weinberg \& Gunn(1990)}]{Weinberg1990}
Weinberg, D.~H. \& Gunn, J.~E. 1990, Mon. Not. R. Astron. Soc., 247, 260

\bibitem[{Zeldovich(1970)}]{Zeldovich1970}
Zeldovich, Y.~B. 1970, Astron. Astrophys., 5, 84

\end{thebibliography}

\end{document}